# Physical Layer Network Coding

Shengli Zhang, Soung-Chang Liew, and Patrick, P. K. Lam

*Abstract* **—A main distinguishing feature of a wireless network compared with a wired network is its broadcast nature, in which the signal transmitted by a node may reach several other nodes, and a node may receive signals from several other nodes simultaneously. Rather than a blessing, this feature is treated more as an interference-inducing nuisance in most wireless networks today (e.g., IEEE 802.11). This paper shows that the concept of network coding can be applied at the physical layer to turn the broadcast property into a capacity-boosting advantage in wireless ad hoc networks. Specifically, we propose a physical-layer network coding (PNC) scheme to coordinate transmissions among nodes. In contrast to "straightforward" network coding which performs coding arithmetic on digital bit streams after they have been received, PNC makes use of the additive nature of simultaneously arriving electromagnetic (EM) waves for equivalent coding operation. And in doing so, PNC can potentially achieve 100% and 50% throughput increases compared with traditional transmission and straightforward network coding, respectively, in multi-hop networks. More specifically, the information-theoretic capacity of PNC is almost double that of traditional transmission in the SNR region of practical interest (higher than 0dB). We believe this is a first paper that ventures into EM-wave-based network coding at the physical layer and demonstrates its potential for boosting network capacity.**

## I. Introduction

ONE of the biggest challenges in the wireless communication research is to deal with the interference at the receiver when signals from multiple sources arrive simultaneously. In the radio channel of the physical layer of wireless networks, data are transmitted through electromagnetic (EM) waves in a broadcast manner. The interference between these EM waves causes the data to be scrambled.

To overcome its negative impact, most schemes attempt to find ways to either reduce or avoid interference through receiver design or transmission scheduling [1]. For example, in 802.11 networks, the carrier-sensing mechanism manages the nodes within the same broadcast range so that at most one source can transmit or receive at any time. This is obviously inefficient when multiple nodes have data to transmit.

While interference causes throughput degradation on wireless networks in general, its negative effect for *multi-hop* ad hoc networks is particularly significant. For example, in 802.11 networks, the theoretical throughput of a multi-hop flow in a linear network is less than 1/4 of the single-hop case due to the "self interference" effect, in which the packet of a hop collides with another packet of a nearby hop [2, 3] for the same traffic flow.

Instead of treating interference as a nuisance to be avoided, we can embrace interference to improve throughput performance with the "right mechanism". To do so in a multi-hop network, the following goals must be met:

1. A relay node must be able to convert simultaneously received signals into interpretable output signals to be relayed to their final destinations.
2. A destination must be able to extract the information addressed to it from the relayed signals.

The capability of network coding to combine and extract information through simple Galois field $GF(2^n)$ additions [4, 5] provides a potential approach to meet such goals. However, network coding arithmetic is generally only applied on bits that have already been correctly received. That is, when the EM waves from multiple sources overlap and mutually interfere, network coding cannot be used to resolve the data at the receiver. So, criterion 1 above cannot be met.

This paper proposes the application of network coding directly within the radio channel at the physical layer. We call this scheme Physical-layer Network Coding (PNC). The main idea of PNC is to create an apparatus similar to that of network coding, but at the physical layer that deals with EM signal reception and modulation. Through a proper modulation-and-demodulation technique at the relay nodes, additions of EM signals can be mapped to $GF(2^n)$ additions of digital bit streams, so that the interference becomes part of the arithmetic operation in network coding.

Historically, Shannon first studied the two-way communication channel, where two nodes simultaneously transmit signals to each other [6]. Recently, two-way relay channel, defined as the bidirectional transmission between two end nodes with relay nodes in between, also begins to attract attention. In [7,8], the channel capacities of both full-duplex and half-duplex two-way relay channels are investigated without the use of network coding. In [9], the authors proposed a transmission scheme that applies network coding in a two-way relay channel. In [10], a scheme based on network coding and channel coding is proposed, and the capacity of a two-way relay channel with network coding is analyzed. The previous work, however, has not considered direct application of network coding at the physical layer. This paper shows that with PNC, the capacity of a two-way relay channel increases by 100% compared to the traditional scheme without network coding, and by 50% compared to straight forward network coding.

The rest of this paper is organized as follows. Section II illustrates how PNC works in a linear three-node multi-hop network and compares its performance with conventional schemes. We show that PNC requires only two time slots for the two end nodes to exchange two frames, one in each direction, via the middle relay node. By contrast, three time slots are needed in straightforward network coding, and four time slots are needed if network coding is not used at all. Section III establishes the general PNC modulation-demodulation mapping principle required to ensure the equivalence of network-coding arithmetic and EM-wave interference arithmetic. Section IV extends the discussion in Section II to a linear *N*-node network consisting of two source/destination nodes at two ends, and *N* - 2 relay nodes in between. We show that PNC can achieve the theoretical upper-bound throughput of the linear network. Section V investigates PNC from an information-theoretic angle. We show that the information-theoretic capacity of PNC is almost double that of the traditional transmission when SNR is higher than 0 dB. Section VI further generalizes PNC application to random networks with multiple source-destination pairs. Section VII concludes this paper.

## II. Illustrating Example: A Three-node Wireless Linear Network

Consider the three-node linear network in Fig. 1. $N_1$ (Node 1) and $N_3$ (Node 3) are nodes that exchange information, but they are out of each other's transmission range. $N_2$ (Node 2) is the relay node between them.

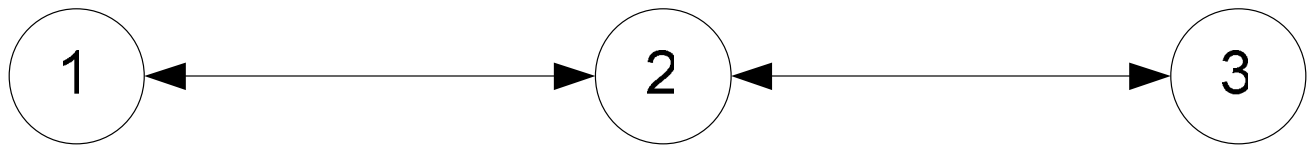

Figure 1. A three-node linear network

This three-node wireless network is a basic unit for cooperative transmission and it has previously been investigated extensively [9, 11-13]. In cooperative transmission, the relay node $N_2$ can choose different transmission strategies, such as Amplify-and-Forward or Decode-and-Forward [11], according to different Signal-to-Noise (SNR) situations. This paper focuses on the Decode-and-Forward strategy. We consider frame-based communication in which a time slot is defined as the time required for the transmission of one fixed-size frame. Each node is equipped with an omni-directional antenna, and the channel is half duplex so that transmission and reception at a particular node must occur in different time slots.

Before introducing the PNC transmission scheme, we first describe the traditional transmission scheduling scheme and the "straightforward" network-coding scheme for mutual exchange of a frame in the three-node network [9, 14].

### A. Traditional Transmission Scheduling Scheme

In traditional networks, interference is usually avoided by prohibiting the overlapping of signals from $N_1$ and $N_3$ to $N_2$ in the same time slot. A possible transmission schedule is given in Fig. 2. Let $S_i$ denote the frame initiated by $N_i$. $N_1$ first sends $S_1$ to $N_2$, and then $N_2$ relays $S_1$ to $N_3$. After that, $N_3$ sends $S_3$ in the reverse direction. A total of four time slots are needed for the exchange of two frames in opposite directions.s

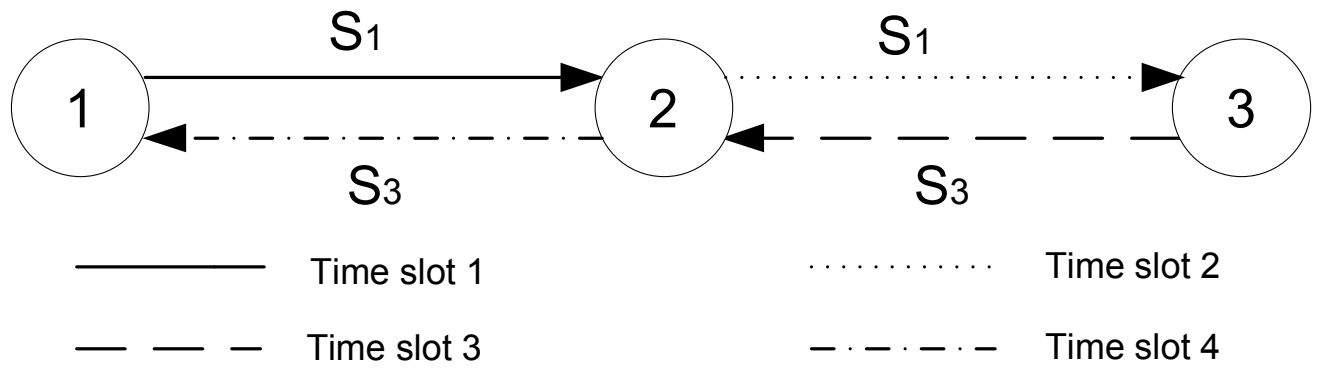

Figure 2. Traditional scheduling scheme

### B. Straightforward Network Coding Scheme

Ref. [14] and [9] outline the straightforward way of applying network coding in the three-node wireless network. Fig. 3 illustrates the idea. First, $N_1$ sends $S_1$ to $N_2$ and then $N_3$ sends frame $S_3$ to $N_2$. After receiving $S_1$ and $S_3$, $N_2$ encodes frame $S_2$ as follows:

$$S_2 = S_1 \oplus S_3 \tag{1}$$

where $\oplus$ denote bitwise exclusive OR operation being applied over the entire frames of $S_1$ and $S_3$. $N_2$ then broadcasts $S_2$ to both $N_1$ and $N_3$. When $N_1$ receives $S_2$, it extracts $S_3$ from $S_2$ using the local information $S_1$, as follows:

$$S_1 \oplus S_2 = S_1 \oplus (S_1 \oplus S_3) = S_3 \tag{2}$$

Similarly, $N_2$ can extract $S_1$. A total of three time slots are needed, for a throughput improvement of 33% over the traditional transmission scheduling scheme.

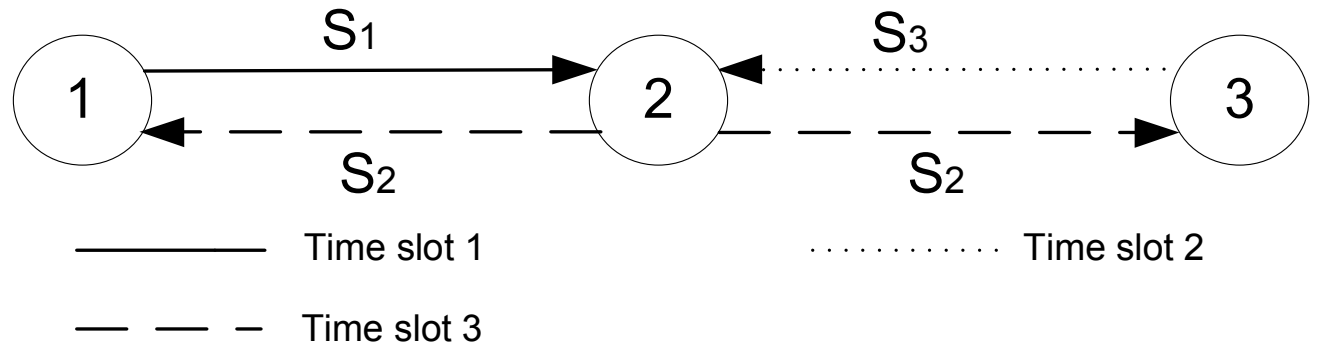

Figure 3. Straightforward network coding scheme

### C. Physical-Layer Network Coding (PNC )

We now introduce PNC. Let us assume the use of QPSK modulation in all the nodes. We further assume symbol-level and carrier-phase synchronization, and the use of power control, so that the frames from $N_1$ and $N_3$ arrive at $N_2$ with the same phase and amplitude. Additional discussions on synchronization issues (i.e., synchronization penalties, what happens when there is no synchronization, etc.) can be found in Appendices 1 and 2. The combined passband signal received by $N_2$ during one symbol period is

$$\begin{aligned} r_2(t) &= s_1(t) + s_3(t) \\ &= [a_1 \cos(\omega t) + b_1 \sin(\omega t)] + [a_3 \cos(\omega t) + b_3 \sin(\omega t)] \\ &= (a_1 + a_3)\cos(\omega t) + (b_1 + b_3)\sin(\omega t) \end{aligned} \tag{3}$$

where $s_i(t)$, $i$ = 1 or 3, is the bandpass signal transmitted by $N_i$ and $r_2(t)$ is the bandpass signal received by $N_2$ during one symbol period; $a_i$ and $b_i$ are the QPSK modulated information bits of $N_i$; and $\omega$ is the carrier frequency. Then, $N_2$ will receive two baseband signals, in-phase ($I$) and quadrature phase ($Q$), as follows:

$$\begin{aligned} I &= a_1 + a_3 \\ Q &= b_1 + b_3 \end{aligned} \quad (4)$$

Note that $N_2$ cannot extract the individual information transmitted by $N_1$ and $N_3$, i.e., $a_1$, $b_1$, $a_3$ and $b_3$, from the combined signal $I$ and $Q$. However, $N_2$ is just a relay node. As long as $N_2$ can transmit the necessary information to $N_1$ and $N_3$ for extraction of $a_1$, $b_1$, $a_3$, $b_3$ over there, the end-to-end delivery of information will be successful. For this, all we need is a special modulation/demodulation mapping scheme, referred to as *PNC mapping* in this paper, to obtain the equivalence of GF(2) summation of bits from $N_1$ and $N_3$ at the physical layer.

Table 1 illustrates the idea of PNC mapping. Recall that a QPSK data stream can be considered as two BPSK data streams: an in-phase stream and a quadrature-phase stream. In Table 1, $s_j^{(I)} \in \{0, 1\}$ is a variable representing the in-phase data bit of $N_j$ and $a_j \in \{-1, 1\}$ is a variable representing the BPSK modulated bit of $s_j^{(I)}$ such that $a_j = 2s_j^{(I)} - 1$. A similar table (not shown here) can also be constructed for the quadrature-phase data by letting $s_j^{(Q)} \in \{0, 1\}$ be the quadrature data bit of $N_j$, and $b_j \in \{-1, 1\}$ be the BPSK modulated bit of $s_j^{(Q)}$ such that $b_j = 2s_j^{(Q)} - 1$.

With reference to Table 1, $N_2$ obtains the information bits:

$$s_2^{(I)} = s_1^{(I)} \oplus s_3^{(I)}; \qquad s_2^{(Q)} = s_1^{(Q)} \oplus s_3^{(Q)} \quad (5)$$

It then transmits

$$s_2(t) = a_2 \cos(\omega t) + b_2 \sin(\omega t) \quad (6)$$

Upon receiving $s_2(t)$, $N_1$ and $N_3$ can derive $s_2^{(I)}$ and $s_2^{(Q)}$ by ordinary QPSK demodulation. The successively derived $s_2^{(I)}$ and $s_2^{(Q)}$ bits within a time slot will then be used to form frame $S_2$. In other words, the operation $S_2 = S_1 \oplus S_3$ in straightforward network coding can now be realized through PNC mapping.

As illustrated in Fig. 4, PNC requires only two time slots for the exchange of one frame (as opposed to three time slots in straightforward network coding and four time slots in traditional scheduling).

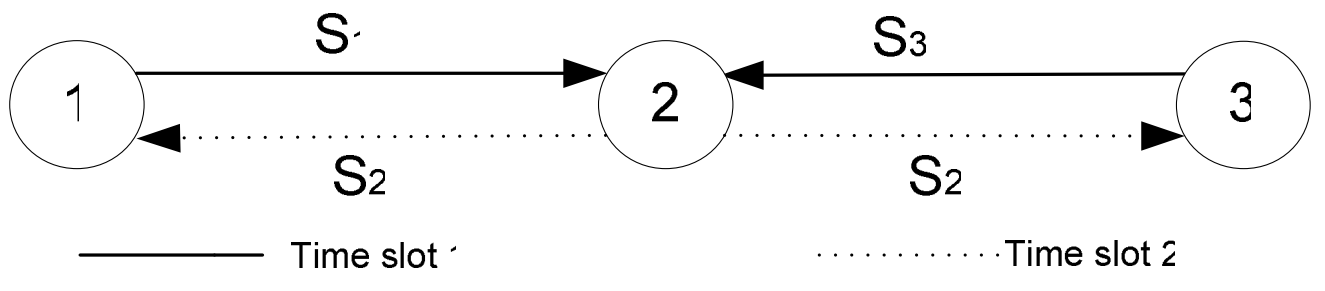

Figure 4. Physical layer network coding

TABLE I. PNC MAPPING: MODULATION MAPPING AT $N_1$, $N_2$; DEMODULATION AND MODULATION MAPPINGS AT $N_3$

| Modulation mapping at $N_1$ and $N_3$, | | | | Demodulation mapping at $N_2$ | | |
|---|---|---|---|---|---|---|
| | | | | Input | Output | |
| Input | | Output | | | Modulation mapping at $N_2$ | |
| | | | | | Input | Output |
| $s_1^{(I)}$ | $s_3^{(I)}$ | $a_1$ | $a_3$ | $a_1 + a_3$ | $s_2^{(I)}$ | $a_2$ |
| 1 | 1 | 1 | 1 | 2 | 0 | -1 |
| 0 | 1 | -1 | 1 | 0 | 1 | 1 |
| 1 | 0 | 1 | -1 | 0 | 1 | 1 |
| 0 | 0 | -1 | -1 | -2 | 0 | -1 |

### *D. BER performance analysis*

We now analyze the bit error rate (BER) performance. Suppose the received signal energy for one bit is unity, and the noise is Gaussian white with density $N_0/2$. For broadcast frames transmitted by $N_2$ to $N_1$ or $N_3$, the BER for all three schemes is simply the standard BPSK modulation $Q(\sqrt{2/N_0})$ [15], where $Q(.)$ is the complementary cumulative distribution function of the zero-mean, unit-variance Gaussian random variable, which is identical to the BER in traditional transmission.

Now, consider reception at $N_2$. Let us first consider straightforward network coding. The ultimate frame is a combination of the two received frames of the same BER $Q(\sqrt{2/N_0})$. Note that the combined XOR bit is in error if and only if the bit from exactly one frame is in error (i.e., having an error in both of the transmitted frames will actually turn out a correct combined frame). Therefore, the BER of the XOR bit of the straightforward network coding is $2Q(\sqrt{2/N_0})(1-Q(\sqrt{2/N_0}))$.

Now, consider PNC. Frames are transmitted by $N_1$ and $N_3$ simultaneously to $N_2$. The BER of the XOR bit can be derived as follows. According to Table 1, the in-phase signal space is {-2, 0, 2} with corresponding probabilities of 25%, 50%, 25% respectively. Note that $a_2$ is only mapped to two integral values: $-1$ when $a_1 + a_3 = 2$ or $a_1 + a_3 = -2$, and 1 when $a_1 + a_3 = 0$. Simply put, $a_2$ is 1 when $a_1 + a_3 = 0$ and is -1 otherwise. Applying the maximum posterior probability criterion [15], we can obtain the optimal thresholds as follows.

$$1=\frac{P(a_1+a_3=0\mid r)}{P(a_1+a_3\neq 0\mid r)}$$
$$=\frac{P(r,a_1+a_3=0)}{P(r,a_1+a_3=2)+P(r,a_1+a_3=-2)}$$
$$=\frac{P(r\mid a_1+a_3=0)P(a_1+a_3=0)}{P(r\mid a_1+a_3=2)P(a_1+a_3=2)+P(r\mid a_1+a_3=-2)P(a_1+a_3=-2)}$$
$$=2\frac{P(r\mid a_1+a_3=0)}{P(r\mid a_1+a_3=2)+P(r\mid a_1+a_3=-2)}$$
$$=2\frac{\exp(-r^2/N_0)/\sqrt{\pi N_0}}{\exp(-(r-2)^2/N_0)/\sqrt{\pi N_0}+\exp(-(r+2)^2/N_0)/\sqrt{\pi N_0}} \quad (7)$$

Solving this equation, we have

$$\gamma_1=-1-\frac{N_0}{4}\ln(1+\sqrt{1-e^{-8/N_0}})$$
$$\gamma_2=1+\frac{N_0}{4}\ln(1+\sqrt{1-e^{-8/N_0}}) \quad (8)$$

where $\gamma_1$ and $\gamma_2$ are the optimal thresholds with which the value of $a_1+a_3$ can be determined as follows. When the received signal amplitude is less than $\gamma_1$ or greater than $\gamma_2$, we declare $a_1+a_3$ to be non-zero (i.e., -2 or 2, respectively) and therefore $a_2$ is -1. Similarly, if the received signal amplitude is greater than $\gamma_1$ and less than $\gamma_2$, we declare $a_1+a_3$ to be zero and therefore $a_2$ is 1. Thus, the BER of the XOR bit can be derived as follows:

$$BER=\frac{1}{2}\int_{-\infty}^{\gamma_1}\frac{1}{\sqrt{\pi N_0}}\exp(\frac{-r^2}{N_0})\,dr+\frac{1}{2}\int_{\gamma_2}^{\infty}\frac{1}{\sqrt{\pi N_0}}\exp(\frac{-r^2}{N_0})\,dr$$
$$+\frac{1}{4}\int_{\gamma_1}^{\gamma_2}\frac{1}{\sqrt{\pi N_0}}\exp(\frac{-(r+2)^2}{N_0})\,dr+\frac{1}{4}\int_{\gamma_1}^{\gamma_2}\frac{1}{\sqrt{\pi N_0}}\exp(\frac{-(r-2)^2}{N_0})\,dr \quad (9)$$

where $r$ is the received in-phase signal at $N_2$. We plot the BER of the XOR bit of PNC, straightforward network coding, and the BER of regular QPSK modulation, in Figure. 5. We can see that the BER of PNC is slightly worse than QPSK, but is slightly better than straightforward network coding. Furthermore, when the SNR is larger than 10dB, the SNR differences between the three schemes is less than 0.1 dB. For simplicity, henceforth, we will ignore these small SNR differences, and assume PNC to have the same BER performance as the traditional 802.11 and straightforward network coding schemes.

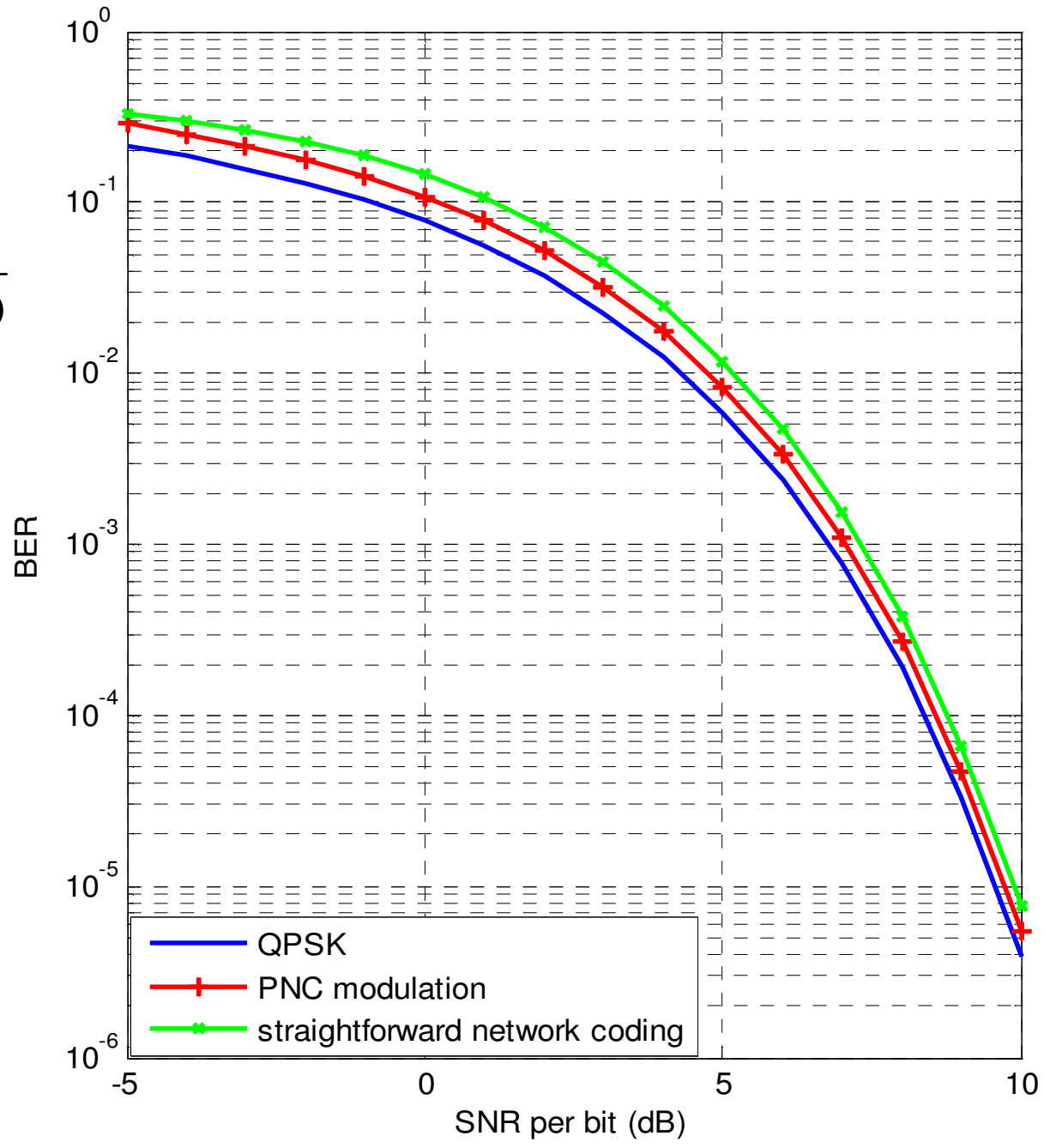

Figure 5. BER at $N_2$ for standard QPSK modulation; BER of the XOR bit at $N_2$ for straightforward network coding and PNC modulation

The last paragraph is related to the BER for the reception at $N_2$. Let us consider the ultimate end-to-end BER at $N_1$ *and* $N_3$. Assume the per-hop BER is small. The end-to-end BER is approximately the summation of the per-hop BER at $N_2$ and the per-hop BER at $N_1$ ($N_3$). Since the per-hop BER from $N_2$ to $N_1$ ($N_3$) is identical for all the schemes, the relation of the end-to-end BER for the three schemes is similar to that in Fig. 5.

For a frame exchange, PNC requires two time slots, 802.11 requires four, while straightforward network coding requires three. Therefore, PNC can improve the system throughput of the three-node wireless network by a factor of 100% and 50% relative to traditional transmission scheduling and straightforward network coding, respectively.

## III. General PNC Modulation-Demodulation Mapping Principle

A specific example of PNC mapping scheme has been constructed in Table 1 for the relay node in a 3-node linear network. We now generalize the PNC mapping principle.

### A. *General PNC Mapping Requirement*

Let us consider the three-node linear network scenario depicted in Fig. 4 again, but now look deeper into its internal operation as shown in Fig. 6. Let $M$ denote the set of digital symbols, and let $\oplus$ be the general binary operation for network-coding arithmetic (note that $\oplus$ is not necessary the bitwise XOR hereinafter). That is, applying $\oplus$ on $m_i, m_j \in M$ gives $m_i \oplus m_j = m_k \in M$. Next, let $E$ denote the set of modulated symbols in the EM-wave domain. Each $m_i \in M$ is mapped to a modulated symbol $e_i \in E$. Let $f: M \rightarrow E$ denote the modulation mapping function such that $f(m_i) = e_i$, $\forall\, m_i$. Note that $f: M \rightarrow E$ is a one-to-one mapping.

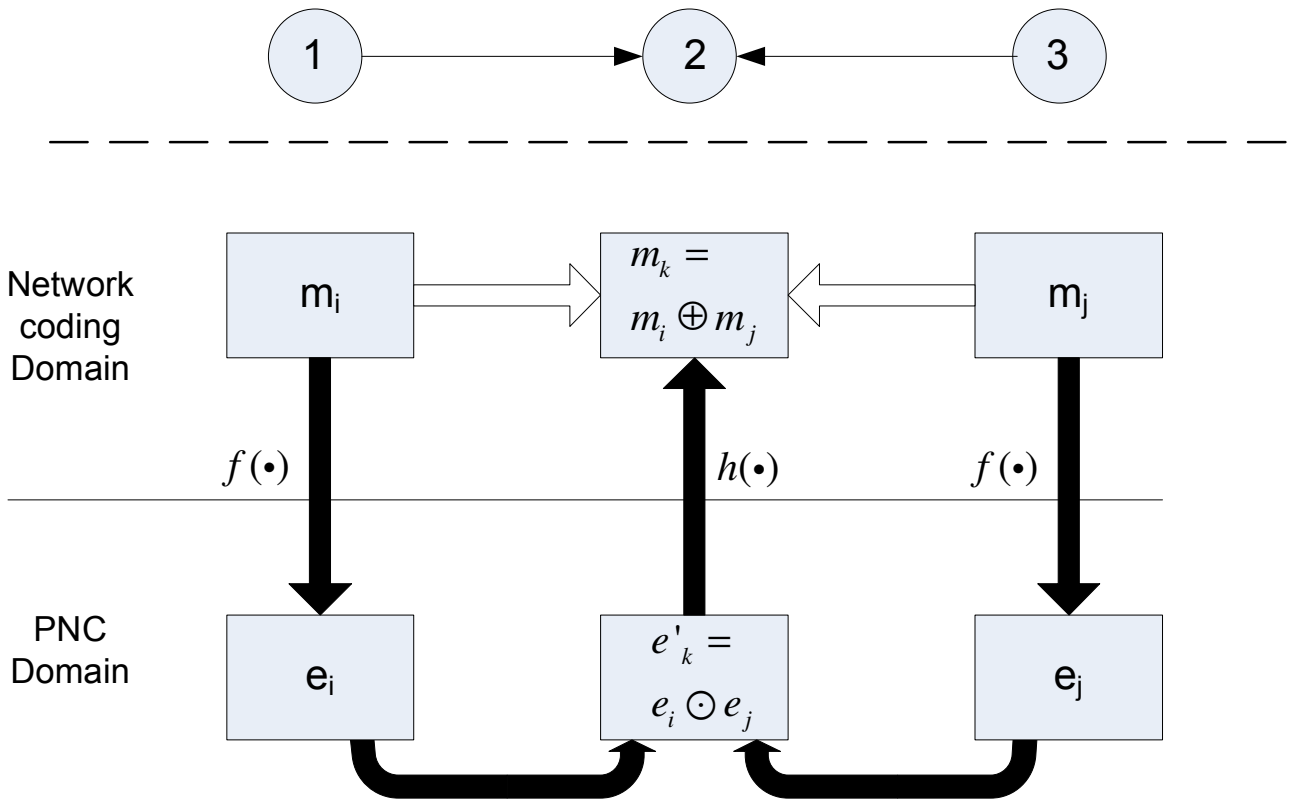

Figure 6. Illustration of PNC mapping

In the EM-wave domain, two signals may combine to yield a composite signal at the receiver. Let $\odot$ represent the binary combination operation. That is, combination of $e_i, e_j \in E$ yields $e_i \odot e_j = e'_k \in E'$, where $E'$ is the domain after the binary operation $\odot$. Note that $E'$ is not the same as $E$ and has a higher cardinality than $E$. For example, for 4-PAM, $E = \{-3, -1, 1, 3\}$, and $E' = \{-6, -4, -2, 0, 2, 4, 6\}$. For BFSK, $E = \{f_1, f_2\}$, and $E' = \{f_1, f_1 \text{ and } f_2, f_2\}$, where $f_1$ and $f_2$ are the constituent frequencies.

Each $e'_k \in E'$ received by the relay node must be mapped to a demodulated symbol $m_k \in M$. Let $h: E' \rightarrow M$ denote the demodulation mapping function such that $h(e'_k) = m_k$. Note that $h: E' \rightarrow M$ is a many-to-one mapping since the cardinality of $E'$ is larger than that of $M$.

To summarize, a PNC transmission scheme consists of the following:

1. Network code specified by $M$ and $\oplus$.
2. One-to-one modulation mapping, $f: M \rightarrow E$.
3. Many-to-one demodulation mapping, $h: E' \rightarrow M$.

Note that while the choices of $M$, $\oplus$, $f: M \rightarrow E$, and $h: E' \rightarrow M$ are up to the network designer, $\odot$ and $E'$ are not because they relate to the fundamental characteristics of EM-wave. Now, there are many possibilities for 1 and 2 above. An interesting question is that, given ($M$, $\oplus$, $f: M \rightarrow E$), whether we can find an appropriate $h: E' \rightarrow M$ to realize PNC. More precisely, for a network code and a modulation scheme, we have the following PNC mapping requirement:

**PNC Mapping Requirement:** Given ($M$, $\oplus$, $f: M \rightarrow E$), there exists $h: E' \rightarrow M$ such that for all $m_i, m_j \in M$, if $m_i \oplus m_j = m_k$, then $h(e_i \odot e_j) = m_k$. That is, $h(f(m_i) \odot f(m_j)) = m_k$.

Fig. 6 illustrates the above requirement, in which the network-coding operation (white arrows) is realized by the PNC operation (dark arrows).

The following proposition specifies the characteristics that the modulation scheme $f: M \rightarrow E$ must possess in order that an appropriate $h: E' \rightarrow M$ can be found.

**Proposition 1:** Consider a modulation mapping $f: M \rightarrow E$. Suppose that $f$ has the characteristic that $e_i \odot e_j = e_p \odot e_q$ implies $m_i \oplus m_j = m_p \oplus m_q$. Then a demodulation mapping $h: E' \rightarrow M$ can be found such that the PNC Mapping Requirement is satisfied. Conversely, if $e_i \odot e_j = e_p \odot e_q$ but $m_i \oplus m_j \neq m_p \oplus m_q$, then $h: E' \rightarrow M$ that satisfies the PNC Mapping Requirement does not exist.

**Proof:** For a given $e'_k \in E'$, one or more pairs of ($e_i$, $e_j$) can be found such that $e_i \odot e_j = e'_k$. If the condition "$e_i \odot e_j = e_p \odot e_q$ implies $m_i \oplus m_j = m_p \oplus m_q$" is satisfied, for any pair of such ($e_i$, $e_j$), $f^{-1}(e_i) \oplus f^{-1}(e_j)$ has the same value as $m_i \oplus m_j$, where $f^{-1}(\cdot)$ is the reverse mapping of the one-to-one mapping $f(\cdot)$. Therefore, $h(e'_k)$ can simply be $f^{-1}(e_i) \oplus f^{-1}(e_j)$ to satisfy the PNC Mapping Requirement.

TABLE II. PNC DEMODULATION SCHEME FOR 4-PAM

| $m_i$ | $m_j$ | $e_i$ | $e_j$ | $e_i + e_j$ | $h(e_i + e_j)$ | $(m_i + m_j) \bmod L$ |
|---|---|---|---|---|---|---|
| 0 | 0 | -3 | -3 | -6 | 0 | 0 |
| 0 | 1 | -3 | -1 | -4 | 1 | 1 |
| 0 | 2 | -3 | 1 | -2 | 2 | 2 |
| 0 | 3 | -3 | 3 | 0 | 3 | 3 |
| 1 | 0 | -1 | -3 | -4 | 1 | 1 |
| 1 | 1 | -1 | -1 | -2 | 2 | 2 |
| 1 | 2 | -1 | 1 | 0 | 3 | 3 |
| 1 | 3 | -1 | 3 | 2 | 0 | 0 |
| 2 | 0 | 1 | -3 | -2 | 2 | 2 |
| 2 | 1 | 1 | -1 | 0 | 3 | 3 |
| 2 | 2 | 1 | 1 | 2 | 0 | 0 |
| 2 | 3 | 1 | 3 | 4 | 1 | 1 |
| 3 | 0 | 3 | -3 | 0 | 3 | 3 |
| 3 | 1 | 3 | -1 | 2 | 0 | 0 |
| 3 | 2 | 3 | 1 | 4 | 1 | 1 |
| 3 | 3 | 3 | 3 | 6 | 2 | 2 |

Conversely, suppose that "$e_i \odot e_j = e_p \odot e_q$ but $m_i \oplus m_j \neq m_p \oplus m_q$". According to the *PNC Modulation-Demodulation Requirement,* the appropriate mapping $h: E' \rightarrow M$ must produce $m_i \oplus m_j = h(e_i \odot e_j) = h(e'_k) = h(e_p \odot e_q) = m_p \oplus m_q$, which contradicts the condition.

### B. PNC for QAM

We now show how Proposition 1 (and its constructive proof for the existence of $h: E' \rightarrow M$) can be used to identify the required PNC mapping in a practical example. Specifically, for $f: M \rightarrow E$, we consider the rectangular M-QAM modulation. QAM can be regarded as the combination of two independent PAM signals, the in-phase signal and quadrature-phase signal. For simplicity, we only consider the in-phase PAM signal here. The analysis for the quadrature phase signal is similar. Suppose the in-phase PAM signal has $L$ levels, so the EM-wave signal space is $E = \{-(L-1), -(L-3), \cdots (L-3), (L-1)\}$. Since the $L$ digital symbols form the set $M = \{0, 1, \cdots (L-2), (L-1)\}$, a possible mapping of $f: M \rightarrow E$ is

$$f(m_i) = e_i = 2m_i - (L-1) \tag{10}$$

Assuming perfect synchronization, the combination of two PAM signals is simply the sum of the magnitudes of the two waves. That is, $e_i \odot e_j = e_i + e_j$.

Suppose the binary network coding operation $\oplus$ is applied on the set $M$ in the following way:

$$m_i \oplus m_j = (m_i + m_j) \bmod L \tag{11}$$

We can now show that $f(\cdot)$ as defined above satisfies the condition in Proposition 1. For any two pairs $(e_i, e_j), (e_p, e_q)$, if $e_i \odot e_j = e_p \odot e_q$, then the corresponding binary network coding result is

$$\begin{aligned} m_i \oplus m_j &= \frac{e_i + L - 1}{2} \oplus \frac{e_j + L - 1}{2} \\ &= \left( \frac{e_i + L - 1}{2} + \frac{e_j + L - 1}{2} \right) \bmod L \\ &= \left( \frac{e_p + L - 1}{2} + \frac{e_q + L - 1}{2} \right) \bmod L = m_p \oplus m_q \end{aligned} \tag{12}$$

because $e_i \odot e_j = e_i + e_j = e_p \odot e_q = e_p + e_q$. Therefore, $e_i \odot e_j = e_p \odot e_q$ implies $m_i \oplus m_j = m_p \oplus m_q$. Based on Proposition 1, an appropriate PNC demodulation mapping exists and can be expressed as follows:

$$\begin{aligned} h(e'_k) &= h(e_i + e_j) = f^{-1}(e_i) \oplus f^{-1}(e_j) \\ &= \left( \frac{e_i + (L-1)}{2} + \frac{e_j + (L-1)}{2} \right) \bmod L \\ &= ((e_i + e_j)/2 - 1) \bmod L \\ &= (e'_k / 2 - 1) \bmod L \end{aligned} \tag{13}$$

## IV. PNC in General Regular Linear Network

In the preceding sections, we have illustrated the basic idea of PNC with a three-node linear wireless network. In this section, we consider the general regular linear network with more than three nodes. For simplicity, we assume the distance between any two adjacent nodes is fixed at $d$.

As will be detailed later, when applying PNC on the general linear network, each node transmits and receives alternately in successive time slots; and when a node transmits, its adjacent nodes receive, and vice versa (see Fig. 7). Let us briefly investigate the signal-to-inference ratio (SIR) given this transmission pattern to make sure that it is not excessive. Consider the worst-case scenario of an infinite chain. We note the following characteristics of PNC from a receiving node's point of view:

1. The interfering nodes are symmetric on both sides.
2. The simultaneous signals received from the two adjacent nodes do not interfere due to the nature of PNC.
3. The nodes that are two hops away are also receiving at the same time, and therefore will not interfere with the node.

Therefore, the two nearest interfering nodes are three hops away. We have the following SIR:

$$SIR = \frac{P_0 / d^\alpha}{2 * \sum_{l=1}^{\infty} P_0 / [(2l+1)d]^\alpha} \tag{14}$$

where $P_0$ is the common transmitting power of nodes and $\alpha$ is the path-loss exponent. Assume the two-ray propagation model where $\alpha = 4$. The resulting SIR is about 16dB and based on Fig. 5, and the impact of the interference on BER is negligible for BPSK. More generally, a thorough treatment should take into account the actual modulation scheme used, the difference between the effects of interference and noise, and whether or not channel coding is used. However, we can conclude that as far as the SIR is concerned, PNC is not worse than *traditional scheduling* (see Section II) when generalized to the *N*-node network. This is because for the generalized traditional scheduling, the interferers are 2, 2, 3, 5, 6, 6, 7, 9, 10, 10,…, hops away and the total interference power is larger than that in the PNC case above. To limit our scope, we obviate the thorough SIR results here.

We now describe the PNC scheme under the general regular linear network more precisely. In the following we first consider the operation of PNC in the simple uni-directional case, followed by the bi-directional case.

### A. Uni-Directional Transmission

Consider a regular linear network with $n$ nodes. Label the nodes as node 1, node 2, …, node $n$, successively with nodes 1 and $n$ being the two source and destination nodes, respectively. Fig. 7 shows a network with $n$ = 5.

Divide the time slots into two types: odd slots and even slots. In the odd time slots, the odd-numbered nodes transmit and the even-numbered nodes receive. In the even time slots, the even-numbered nodes transmit and the odd-numbered nodes receive. Suppose that node 1 is to transmit frames $X_1$, $X_2$, …. to the destination node $n$.

Fig. 7 shows the sequence of frames being transmitted by the nodes in a 5-node network. In slot 1, node 1 transmits $X_1$ to node

2. In slot 2, node 2 transmits $X_1$ to node 3; node 2 also stores a copy of $X_1$ in its buffer. In slot 3, node 1 transmits $X_2$ to node 2, and node 3 transmits $X_1$ to node 4, but the transmission also reaches node 2; node 3 stores a copy of $X_1$ in its buffer. Thus, node 2 receives $X_1 \oplus X_2$. Node 2 then "adds" the inverse of its stored copy of $X_1$, $X_1^{-1}$, to $X_1 \oplus X_2$ to obtain $X_1^{-1} \oplus X_1 \oplus X_2 = X_2$. In slot 4, node 2 transmits $X_2$ and node 4 transmits $X_1$. In this way, node 5 receives a copy of $X_1$ in slot 4. Also, in slot 4, node 3 receives $X_1 \oplus X_2$ and then use $X_1^{-1}$ to obtain $X_1^{-1} \oplus X_2 \oplus X_1 = X_2$.

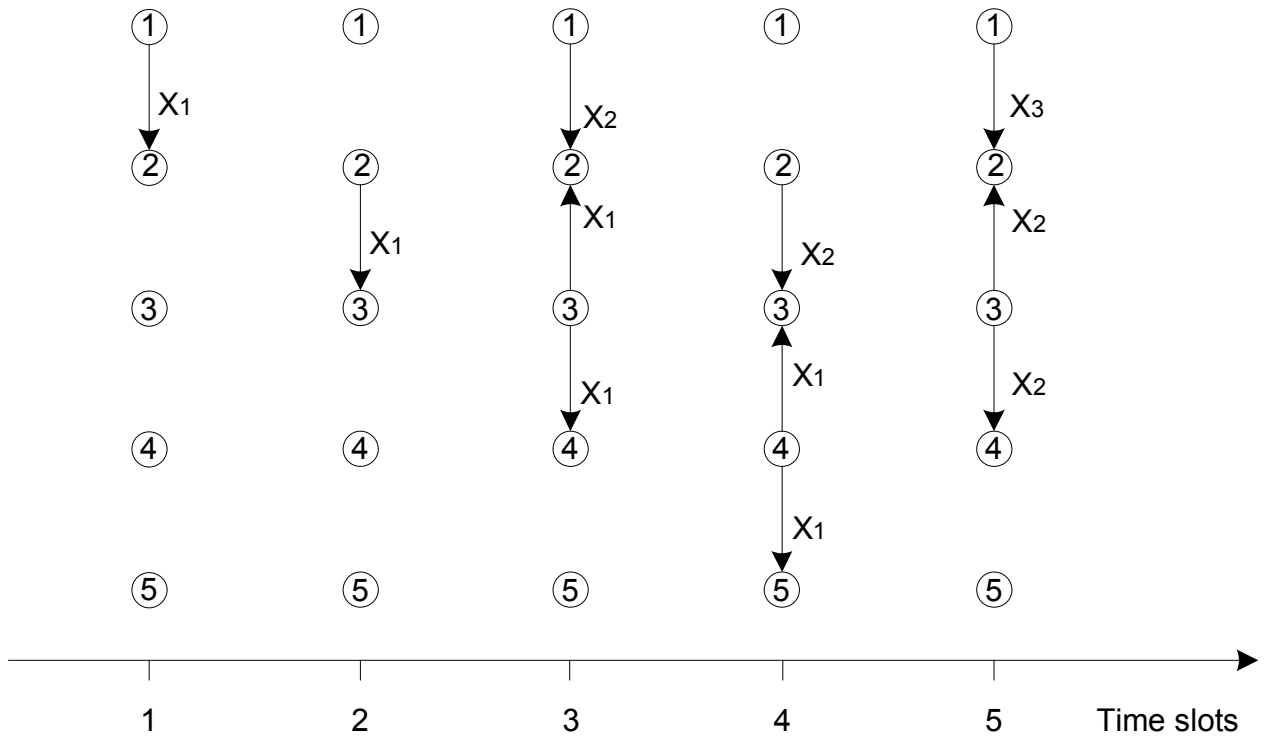

Figure 7. Uni-directional PNC transmission in linear network

**Theorem 1:** For the regular linear network, PNC can achieve the upper-bound throughput, 0.5 frame/time slot, for uni-directional transmission from one end of the network to the other end.

*Proof:* In a multi-hop transmission, each half-duplex relay node must use one time slot to receive a frame and another to send it out. So, it can at most relay one frame in two time slots (i.e., the upper bound is 0.5 frame/time slot). On the other hand, in PNC, each relay node transmits and receives frames in alternative time slot with no idle time, and it relay information contained in a frame in every two time slots. So, it achieves this upper-bound throughput.

A point of interest is that for uni-directional transmission, straightforward network coding does not have an advantage over the traditional scheduling scheme (e.g., nodes 1 and 3 still cannot transmit together because of the "collision" at node 2) and they both have throughput of 1/3 frame per time slot; on the other hand, PNC does.

### *B. PNC for Bi-directional Transmission*

Let us now consider the situation when the two end nodes (i.e., nodes 1 and $n$) transmit frames to each other with the same rate via multiple relay nodes. Suppose that node 1 is to transmit frames $X_1$, $X_2$, …. to node $n$, and node $n$ is to transmit frames $Y_1$, $Y_2$, …. to node 1.

Fig. 8 shows the sequence of frames being transmitted by the nodes in a 5-node network. As in the uni-directional case, a relay node stores a copy of the frame it sends in its buffer. It "adds" the inverse of this stored frame to the frames that it receives from the adjacent nodes in the next time slot to retrieve the "new information" being forwarded by either side. With reference to Fig. 8, we see that a relay node forwards two frames, one in each direction, every two time slots. So, the throughput is 0.5 frame/time slot in each direction.

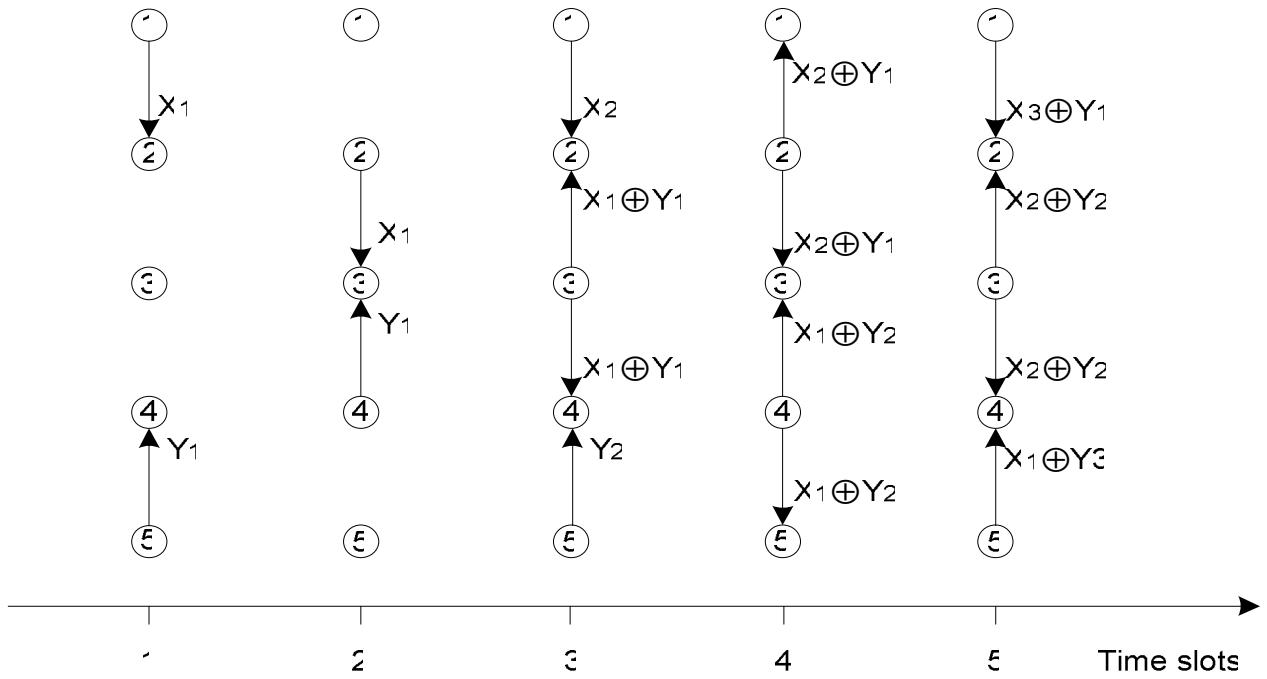

Figure 8. Bi-direction PNC transmission in linear network

**Theorem 2:** For the regular linear network, PNC can achieve the upper-bound throughput, 0.5 frame/time slot in each direction, for bi-directional transmissions between two end nodes.

*Proof*: If the rates from both sources are identical, the proof is similar to the one given for Theorem 1. In general, let us denote the data rate in one direction by $V_X$ and the data rate in another direction by $V_Y$. First, we note that it is simply not feasible for either $V_X$ or $V_Y$ to exceed 0.5 frame/slot because it would exceed the capability of the half-duplex channel. Define the slacks as $S_X = 0.5 - V_X$, and $S_Y = 0.5 - V_Y$. We insert dummy null frames $\varnothing$ into the buffers at nodes 1 and $n$, so that nothing is transmitted during a slot when only a null frame comes up in the buffers (more detailed discussion of null frame can be found in the section, "formal description of PNC frame-forwarding mechanism", below). The rate at which null frames appear correspond to the slacks $S_X$ and $S_Y$. So, essentially the transmission rates are $V_X$ and $V_Y$.

In the regular linear network, if all the frames to be delivered are already available at the sources at the inception of the transmission, there is no incentive to use rates lower than 0.5. Rates smaller than 0.5 is relevant in two situations: 1) the source node generates frames in real-time at a rate smaller than 0.5; 2) a link between two nodes is used by many bi-directional PNC flows. The latter is particularly relevant in a general network topology, in which the per-directional link throughput has to be shared among all the flows that traverse the link.

### *C. Formal Description of PNC Frame-Forwarding Mechanism*

This section may be skipped without sacrificing continuity. The time slots are divided into odd and even slots, and during odd slots, odd nodes transmit and during even slots, even nodes transmit. For generality, we allow for the possibility of a null frame, denoted by $\varnothing$ (this is relevant to the proof of Theorem 2 above and also for throughput allocation in a general network with many PNC flows). When we say an odd (even) node transmits a null frame in an odd (even) slot, we mean the node keeps silence and transmits nothing; similarly, when we say an odd (even) node receives a null frame in an even (odd) slot, we mean the node receives nothing. The null frame has the following property:

$$\begin{aligned} & X_l \oplus \varnothing = X_l \quad \text{for all } X_l \\ & X_l \oplus X_l^{-1} = \varnothing \quad \text{for all } X_l \\ & \varnothing^{-1} = \varnothing \end{aligned} \tag{15}$$

In terms of protocol implementation, if a transmitter intends to keep silence during one of its assigned transmission time slots, it should inform its two adjacent receivers at the beginning of the time slot, so that the receivers can revert back to ordinary non-PNC demodulation scheme to effect the above operational outcome. There is no need to inform the adjacent nodes during a reception (unassigned) slot of a node because it is understood that nothing will be transmitted by the node.

We now give the formal description of the PNC frame-forwarding mechanism for a general situation. The data rates in the two directions are not necessarily the same in this general scheme. We assume that each node $i$ has a buffer $B_i$ containing alternately the frame "to be transmitted" and the frame "just transmitted" by node $i$ in successfully time slots. Initially, $B_i$ is empty for all $i$. Let $S_i[j]$ and $R_i[j]$ denote the frames transmitted and received by node $i$ in the time slot $j$, respectively. Let $B_i[j]$ be the buffer content of $B_i$ in time slot $j$. Assuming the transmissions start in time slot 1, we have the following initial condition for node $i$:

$$\begin{aligned} & S_i[j] = R_i[j] = B_i[j] = \varnothing, \quad j \le 0, \forall i \\ & X_l = Y_l = \varnothing \quad l \le 0 \end{aligned} \tag{16}$$

Without loss of generality, let us assume that $n$ is odd. The case of even $n$ can be easily extrapolated from the same procedure presented here. The following equations describe the operation at node 1:

$$S_1[j] = \begin{cases} B_1[j] & \text{for } j = 1,3,5,\cdots \\ \varnothing & \text{for } j = 2,4,6,... \end{cases}$$

$$R_1[j] = \begin{cases} \varnothing & \text{for } j = 1,3,5,\cdots \\ S_2[j] & \text{for } j = 2,4,6,... \end{cases}$$

$$B_1[j] = \begin{cases} X_{(j+2)/2} \oplus B[j-1] & \text{for } j = 1,3,5,\cdots \\ X_{j/2}^{-1} \oplus R_1[j] & \text{for } j = 2,4,6,... \end{cases} \tag{17}$$

The following equations describe the similar operation at node $n$:

$$S_n[j] = \begin{cases} B_n[j] & \text{for } j = 1,3,5,\cdots \\ \varnothing & \text{for } j = 2,4,6,... \end{cases}$$

$$R_n[j] = \begin{cases} \varnothing & \text{for } j = 1,3,5,\cdots \\ S_{n-1}[j] & \text{for } j = 2,4,6,... \end{cases} \tag{18}$$

$$B_n[j] = \begin{cases} Y_{(j+1)/2} \oplus B_n[j-1] & \text{for } j = 1,3,5,\cdots \\ Y_{j/2}^{-1} \oplus R_n[j] & \text{for } j = 2,4,6,... \end{cases}$$

For odd nodes $i \in \{3,5,...,n-2\}$, we have

$$S_i[j] = \begin{cases} B_i[j] & \text{for } j = 1,3,5,\cdots \\ \varnothing & \text{for } j = 2,4,6,... \end{cases}$$

$$R_i[j] = \begin{cases} \varnothing & \text{for } j = 1,3,5,\cdots \\ S_{i-1}[j] \oplus S_{i+1}[j] & \text{for } j = 2,4,6,... \end{cases} \tag{19}$$

$$B_i[j] = \begin{cases} B_i[j-1] & \text{for } j = 1,3,5,\cdots \\ B_i^{-1}[j-1] \oplus R_i[j] & \text{for } j = 2,4,6,... \end{cases}$$

For even nodes $i \in \{2,4,...,n-1\}$, we have

$$S_i[j] = \begin{cases} \varnothing & \text{for } j = 1,3,5,\cdots \\ B_i[j] & \text{for } j = 2,4,6,... \end{cases}$$

$$R_i[j] = \begin{cases} S_{i-1}[j] \oplus S_{i+1}[j] & \text{for } j = 1,3,5,\cdots \\ \varnothing & \text{for } j = 2,4,6,... \end{cases} \tag{20}$$

$$B_i[j] = \begin{cases} B_i^{-1}[j-1] \oplus R_i[j] & \text{for } j = 1,3,5,\cdots \\ B_i[j-1] & \text{for } j = 2,4,6,... \end{cases}$$

It can be shown from the above that $B_1[j] = Y_{(j-n+3)/2}$, and $B_n[j] = X_{(j-n+3)/2}$, for $j = 2,4,6,...$ That is, after some delay, the information from one end reaches the other end and can be decoded there based on the above procedure.

## V. Two-way Relay Channel Capacity with PNC Scheme

In previous sections, we have given an overview of PNC, and derived its throughput performance from the upper layer's (i.e., the link layer's) point of view. In this section, we analyze the information-theoretic capacity of the two-way relay channel of the three-node network under PNC from the physical layer's point of view. We compare the PNC result with the traditional and the straightforward network coding schemes.

The discussion of this section is based on the following assumptions. At the transmitters, the signals are again modulated with BPSK (note that the performance of BPSK is the same as QPSK) and are transmitted with unit power. At the receivers, we assume additive Gaussian noise with all nodes having identical noise variance. Hard decision demodulation is assumed prior to channel decoding. For simplicity, hard decision is regarded as part of the channel for our computation of the channel capacity. We also assume that both end nodes have the same amount of data to transmit, and the capacity, unlike the previous sections, is defined as the number of bits exchanged during one transmission cycle. Specifically, the half-duplex channel is time-divided into "upstream" and "downstream" sub-channels within a transmission cycle during which the end nodes and the relay node transmit, respectively.

### A. Capacity of Traditional Transmission Scheme

For the traditional transmission scheme, a transmission cycle is divided into four time-divided sub-channels for information exchange. From the information theoretical view point, the four sub-channels are identical and each of them can be represented by

$$r = a + n \tag{21}$$

where, $a \in \{1, -1\}$ is the input of the channel, $r$ is the output of the channel and $n$ is the Gaussian noise. With coherent demodulation and hard decision, the Gaussian channel becomes a standard Binary Symmetric Channel (BSC) with output

$$y = s \oplus e \tag{22}$$

where $e \in \{0,1\}$ is the error pattern. The crossover probability is $p(e=1) = Q(\sqrt{2/N_0})$. The capacity of the BSC channel is [6]

$$C = 1 - H(p) \tag{23}$$

where, $H(p) = -p\log p - (1-p)\log(1-p)$ is the entropy of the binary distribution with probability $p$. Obviously, the durations of the time-divided sub-channels should be identical, which is 1/4. The effective capacity of the traditional transmission scheme is therefore

$$C_1 = C/4 = (1 - H(p))/4 \tag{24}$$

## B. Capacity of Straightforward Network Coding Scheme

In straightforward network coding, the transmission cycle is divided into three equal time slots. And each sub-channel is a BSC channel given by (23). Obviously, the effective capacity of this scheme is

$$C_2 = C/3 = (1 - H(p))/3 \tag{25}$$

## C. Capacity of Physical Network Coding Scheme

In PNC, there are two time slots per transmission cycle: one for multiple-access from the neighboring nodes to the relay node, and the other for broadcast from the relay node to the neighboring nodes. The broadcast channel is simply the BSC given by (23). The analysis of the capacity of the multiple-access channel is more complicated.

For the multiple access capacity, let us first look at the multiple access diagram shown in Fig. 9, where $s_1$, $s_3$ are the information bits of node $N_1$ and $N_3$ respectively; $a_1$ and $a_3$ are the symbols after channel encoding and BPSK modulation. (In the following sections, we also use $s_i$ to denote the information block to be encoded by $N_i$ and $a_i$ to denote the codeword of $s_i$.) After channel encoding and modulation, $a_1$ and $a_3$ are sent to the relay node $N_2$ synchronously. Since $a_1$ and $a_3$ are independent BPSK symbols, we can choose uniform distribution for $a_1, a_3$, which would give the optimal channel capacity. Then, the distribution of the input $a_1 + a_3$ to the multiple access channel is

$$\begin{aligned} &p(a_1 + a_3 = 2) = p(a_1 + a_3 = -2) = 1/4 \\ &p(a_1 + a_3 = 0) = 1/2 \end{aligned} \tag{26}$$

The output of the multiple-access channel can be represented by

$$r = a_1 + a_3 + n \tag{27}$$

After hard decision, we obtain the estimation of the input $a_1 + a_3$, denoted by $\widehat{a_1 + a_3}$.

The multiple access procedure of PNC is different from the traditional point-to-point transmission. For the transmitters, the inputs of the two separated encoders are independent. For the receiver, the input of the decoder is the summation of the two coded signals and the target output of the decoder is the exclusive OR of the original information. As a result, current channel coding schemes for point-to-point transmissions are sub-optimal for this system. The capacity is also unknown. In the following sections, we give an upper bound and a lower bound on the PNC capacity.

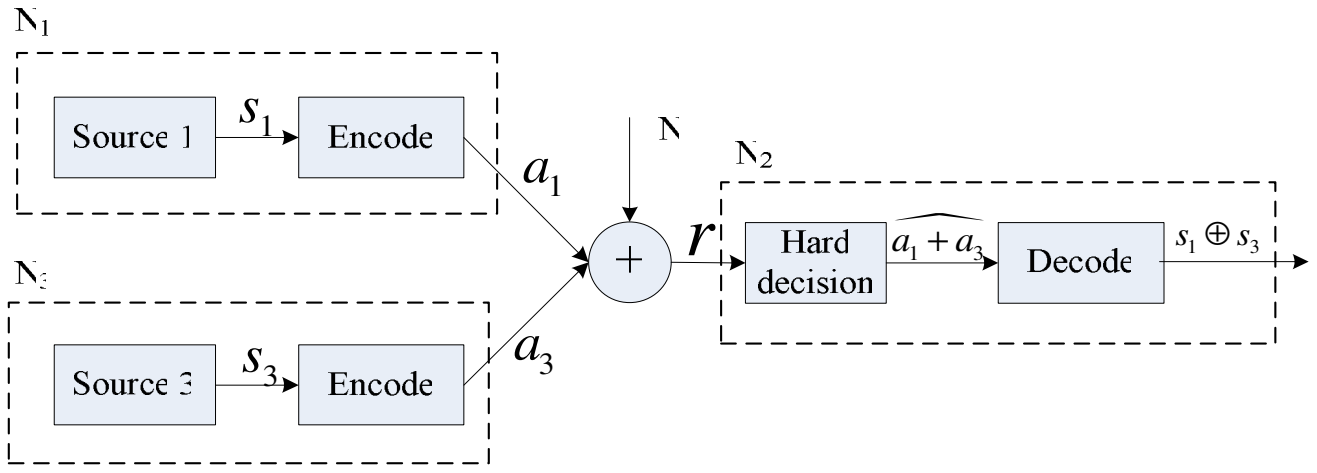

Figure 9. Multiple access diagram for PNC system

### 1) Upper bound of PNC capacity:

For an upper bound, let us decompose the PNC multiple-access channel into a serial concatenation of two sub-channels as shown in Fig. 10. The first sub-channel corresponds to the superposition of the two input signals; the second sub-channel is a channel with additive Gaussian noise and hard decision. Obviously, the capacity of the whole channel will not exceed that of either sub-channel. And we take the capacity of the second sub-channel as the upper bound of this multiple access channel.

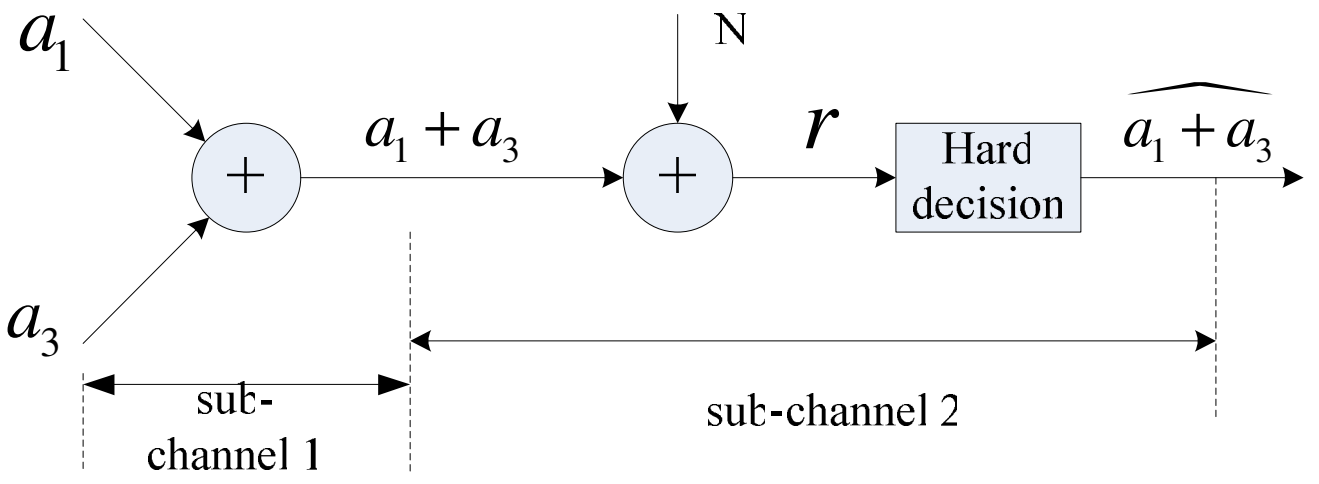

Figure 10. Channel decomposition of PNC multi-access procedure.

With the optimal decision threshold given in (8), the received symbol is then $\widehat{a_1 + a_3} \in \{-2, 0, 2\}$. The mapping of the channel is shown in Fig. 11, where the symbols on the left hand side represent the input of the channel (i.e., $a_1 + a_3$), while the symbols on the right hand side represent the output of the channel (i.e., $\widehat{a_1 + a_3}$) after hard decision. Suppose $p_1$, $p_2$ and $p_3$ are the crossover probabilities from a symbol in $a_1 + a_3$ to a symbol in $\widehat{a_1 + a_3}$ such that

$$\begin{aligned} p_1 &= P(\widehat{a_1 + a_3} = 2 \mid a_1 + a_3 = 0) = P(\widehat{a_1 + a_3} = -2 \mid a_1 + a_3 = 0) \\ p_2 &= P(\widehat{a_1 + a_3} = 0 \mid a_1 + a_3 = 2) = P(\widehat{a_1 + a_3} = 0 \mid a_1 + a_3 = -2) \\ p_3 &= P(\widehat{a_1 + a_3} = -2 \mid a_1 + a_3 = 2) = P(\widehat{a_1 + a_3} = 2 \mid a_1 + a_3 = -2) \end{aligned} \tag{28}$$

The transition probabilities from each symbol of $a_1 + a_3$ to each symbol of $\widehat{a_1 + a_3}$ can then be deduced.

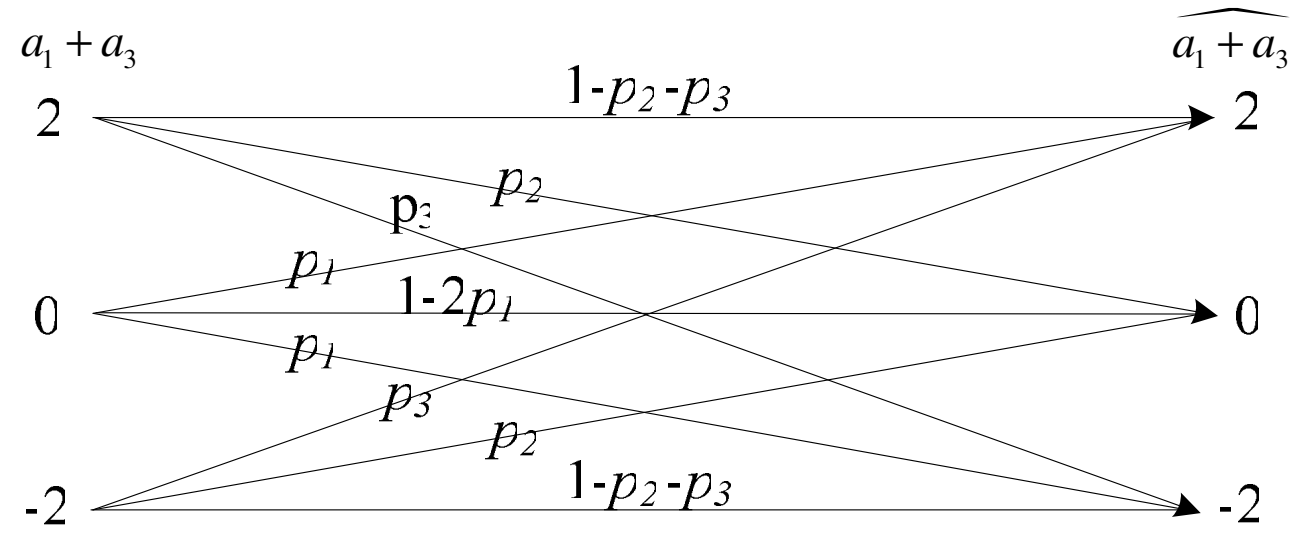

Figure 11. Channel mapping of PNC (based on Fig. 10)

According to the decision rule in (8), we have:

$$
\begin{aligned}
&p_1 = Q(\gamma_2\sqrt{2/N_0}) \\
&p_2 = \begin{cases} 1 - Q((\gamma_2 - 2)\sqrt{2/N_0}) - Q((\gamma_2 + 2)\sqrt{2/N_0}) & \gamma_2 > 2 \\ Q((2-\gamma_2)\sqrt{2/N_0}) - Q((\gamma_2 + 2)\sqrt{2/N_0}) & 1 < \gamma_2 \le 2 \end{cases} \\
&p_3 = Q((\gamma_2 + 2)\sqrt{2/N_0})
\end{aligned} \tag{29}
$$

Based on the distribution of $a_1 + a_3$ given in (26) and the mapping given in Fig. 11, the distribution of $\widehat{a_1 + a_3}$ is

$$
\begin{aligned}
&p(\widehat{a_1 + a_3} = 2) = p(\widehat{a_1 + a_3} = -2) = (1 - p_2 + 2p_1)/4 \\
&p(\widehat{a_1 + a_3} = 0) = (1 + p_2 - 2p_1)/2
\end{aligned} \tag{30}
$$

The mutual information of such a channel, the upper bound of the multiple-access channel capacity, can be calculated as follows:

$$
\begin{aligned}
C' &= I(a_1 + a_3; \widehat{a_1 + a_3}) = H(\widehat{a_1 + a_3}) - H(\widehat{a_1 + a_3} \mid a_1 + a_3) \\
&= H(\widehat{a_1 + a_3}) - \frac{1}{2}[H(p_2, p_3) + H(p_1, p_1)] \\
&= H(\frac{1 - p_2 + 2p_1}{4}, \frac{1 - p_2 + 2p_1}{4}) - \frac{1}{2}[H(p_2, p_3) + H(p_1, p_1)]
\end{aligned} \tag{31}
$$

where $H(p, q) = -p\log p - q\log q - (1-p-q)\log(1-p-q)$ .

The relay node $N_2$ can at most receive $C'$ bits of information successfully. $N_2$ will then map the received symbols to '0' or '1' according to Table I. And the distribution is (see Fig. 11)

$$
\begin{aligned}
&p(s_2 = 1) = (1 - p_2 + 2p_1)/2 \\
&p(s_2 = 0) = (1 + p_2 - 2p_1)/2
\end{aligned} \tag{32}
$$

Note that there is some information loss due to this operation which corresponds to the mapping from both '2' and '-2' to the single bits '0'. The upper bound of the effective capacity for this multiple access channel is therefore

$$
\begin{aligned}
C_m^{up} &= C'\frac{H(s_2)}{H(\hat{a})} = \\
&H(\frac{1 - p_2 + 2p_1}{2})[1 - \frac{H(p_2, p_3) + H(p_1, p_1)}{2H(\frac{1 - p_2 + 2p_1}{4}, \frac{1 - p_2 + 2p_1}{4})}]
\end{aligned} \tag{33}
$$

Since the broadcast channel is BSC, its capacity is again

$$C_b = C = 1 - H(p) \tag{34}$$

The upper bound of the overall capacity of PNC is therefore

$$C_3^{up} = \frac{1}{1/C_m^{up} + 1/C_b} \tag{35}$$

*2) Lower bound of PNC capacity:*

For the lower bound of the capacity, we propose a feasible channel coding scheme of PNC. With regard to Fig. 4, suppose an identical linear channel coding method, denoted by $\Gamma$ , is used at both $N_1$ and $N_3$ . We have

$$a_1 = \Gamma(s_1) \quad \text{and} \quad a_3 = \Gamma(s_3) \tag{36}$$

$$\Gamma(s_1 \oplus s_3) = \Gamma(s_1) \oplus \Gamma(s_3) = a_1 \oplus a_3 \tag{37}$$

The received signal at node $N_2$ is $r = a_1 + a_3 + n$ . After hard decision and PNC mapping given in TABLE 1, we get $r_2 = \widehat{a_1 \oplus a_3} \in \{-1, 1\}$ . Obviously, the original system is equivalent to a virtual system as shown in Fig. 12. Specifically, channel-encode $s_1$ and $s_2$ as in the real system is the same as channel-encode $s_1 \oplus s_3$ jointly as in the virtual system thanks to the linearity of $\Gamma(\cdot)$ (see (37)).

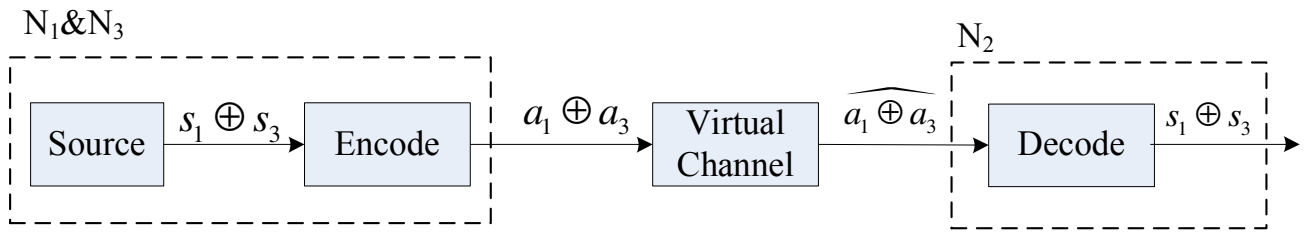

Figure 12. Equivalent system of PNC transmission

It is well known that some linear codes, e.g., Turbo code [16] and LDPC code [17], can approach the Shannon capacity asymptotically. Thus, we can conclude that the Shannon capacity of the virtual channel shown in Fig. 12 is achievable. Therefore, it can well serve as the lower bound of the PNC capacity.

We now calculate the capacity of the virtual channel as follows. With the optimal decision threshold, we can obtain the conditional probability of the channel input and output:

$$
\begin{aligned}
&p(\widehat{a_1 \oplus a_3} = 1 \mid a_1 \oplus a_3 = 1) \\
&= p(\widehat{a_1 + a_3} = 0 \mid a_1 + a_3 = 0) = 1 - 2p_1
\end{aligned} \tag{38}
$$

$$
\begin{aligned}
&p(\widehat{a_1 \oplus a_3} = 1 \mid a_1 \oplus a_3 = 0) \\
&= p(\widehat{a_1 + a_3} = 0 \mid a_1 + a_3 = \pm 2) = p_2
\end{aligned} \tag{39}
$$

$$
\begin{aligned}
&p(\widehat{a_1 \oplus a_3} = 0 \mid a_1 \oplus a_3 = 1) \\
&= p(\widehat{a_1 + a_3} = \pm 2 \mid a_1 + a_3 = 0) = 2p_1
\end{aligned} \tag{40}
$$

$$
\begin{aligned}
&p(\widehat{a_1 \oplus a_3} = 0 \mid a_1 \oplus a_3 = 0) \\
&= p(\widehat{a_1 + a_3} = \pm 2 \mid a_1 + a_3 = \pm 2) = 1 - p_2
\end{aligned} \tag{41}
$$

Based on (38-41), we can obtain the mapping of the virtual channel as shown in Fig. 13. The probabilities $p_1$ and $p_2$ shown in Fig. 13 are identical to the ones given in (28).

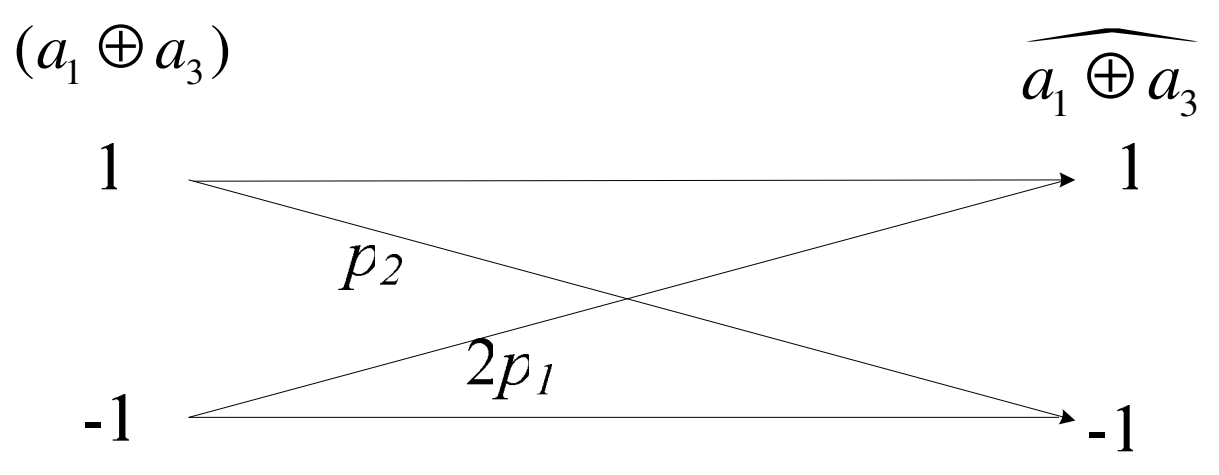

Figure 13. mapping of the virtual channel (based on Fig. 12)

The input distribution of the virtual channel is uniform. And the mutual information (i.e., the lower bound of PNC multiple access channel capacity) is

$$\begin{aligned} C_m^{lo} &= I(a_1 \oplus a_3; \widehat{a_1 \oplus a_3}) \\ &= H(\widehat{a_1 \oplus a_3}) - H(\widehat{a_1 \oplus a_3} \mid a_1 \oplus a_3) \\ &= H((1-2p_1+p_2)/2) - [H(\widehat{a_1 \oplus a_3} \mid 1) - H(\widehat{a_1 \oplus a_3} \mid 0)]/2 \\ &= H((1-2p_1+p_2)/2) - H(2p_1)/2 - H(p_2)/2 \end{aligned} \quad (42)$$

The lower bound of the overall capacity of PNC is therefore

$$C_3^{lo} = \frac{1}{1/C_m^{lo} + 1/C_b} \quad (43)$$

where $C_b$ is obtained from (34).

### D. *Capacity comparison:*

The capacities of the three schemes versus SNR are shown in the Fig. 14. The lower bound capacity of the PNC is better than the other two schemes when SNR is higher than -5dB. And the upper bound capacity of PNC is always better than the other two. TABLE.III shows the capacity gains of PNC and straightforward network coding with respect to traditional transmission. As SNR increases, the gain of the PNC upper bound capacity decreases and the gain of the PNC lower bound capacity increases. When the SNR is higher than 5dB, the two bounds of the PNC capacity converge, and the capacity gain approaches 3/2 and 2 with respect to straightforward network coding and traditional scheme. This gain is determined by the saved time slots considered in section II when SNR is high.

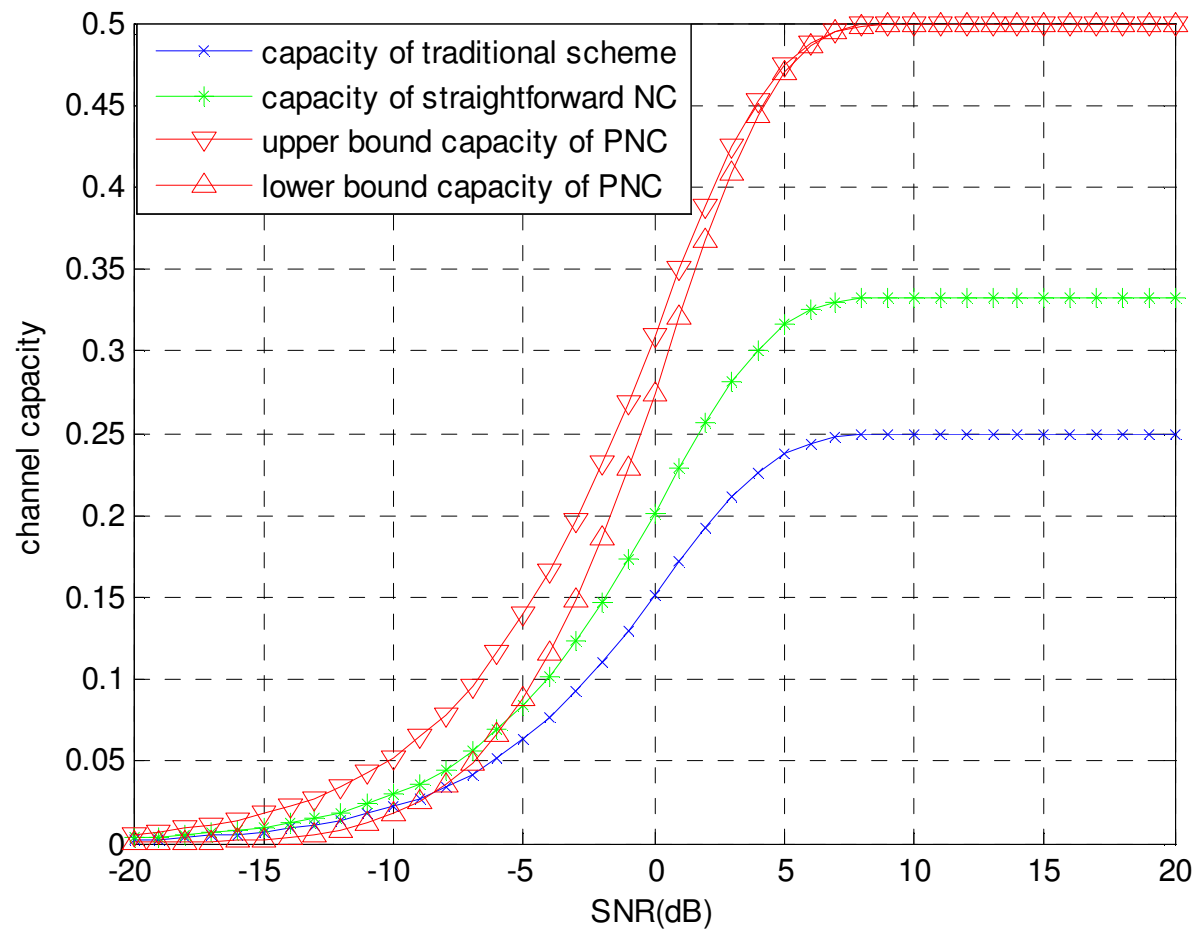

Figure 14. Capacity Comparison

TABLE III. CAPACITY GAIN UNDER DIFFERENT SNRS

| SNR | -20 | -15 | -10 | -5 | 0 | 5 | 10 | 15 | 20 |
|---|---|---|---|---|---|---|---|---|---|
| PNC upper bound gain | 2.51 | 2.48 | 2.39 | 2.21 | 2.05 | 2.00 | 2.00 | 2.00 | 2.00 |
| PNC lower bound gain | 0.13 | 0.35 | 0.82 | 1.41 | 1.82 | 1.98 | 2.00 | 2.00 | 2.00 |
| Straightforward NC gain | 4/3 | 4/3 | 4/3 | 4/3 | 4/3 | 4/3 | 4/3 | 4/3 | 4/3 |

Although the capacity derived here is that of the three-node case with one relay node, the method is also valid for the cases where there are multiple relay nodes between the two end nodes.

## VI. RESOURCE ALLOCATION WITH PNC: AN ARCHITECTURAL OUTLINE

Our discussions so far has only focused on a single flow. We briefly outline a possible architecture for using PNC to support multiple flows in a general network in this section.

### A. *Partitioning of Time Resources*

By nature, PNC is suitable for flows with bidirectional isochronous traffic with implied rate requirements; it is not as suitable for uni-directional best-effort flows. Based on this observation, we can divide time into periodically repeating intervals. Within each interval, there are two subinterval. The first subinterval is dedicated to PNC traffic and the second subinterval is dedicated to non-PNC traffic. The second subinterval may contain best-effort traffic as well as isochronous traffic that does not make use of PNC. The first subinterval, however, contains only PNC isochronous traffic.

Each PNC flow passing through a node is dedicated specific time slots within the first subinterval. In the parlance of the previous discussion, an "odd" node will only transmit data frames in the "odd" time slots within the first subinterval. With multiple PNC flows, the odd time slots are further partitioned so that different PNC flows will use different odd time slots.

The relative lengths of the first and second subintervals can be adjusted dynamically based on the traffic demands and the relative portions of the isochronous traffic that can exploit PNC. Some isochronous flows passing through a node cannot make use of PNC. This will be the case, for example, when the end-to-end path of a flow consists of several PNC chains, in between of which the conventional multi-hop scheme is used (see Part B of this section for further details). It may be necessary to break a long end-to-end path into multiple PNC chains to simplify resource management as well as to limit the synchronization overhead (see Appendixes 1 and 2 for discussions on synchronization overhead). The conventional multi-hop scheme is also needed in portions of the network in which PNC is not possible due to physical constraints.

Conceptually, the rates of the isochronous traffic can be described by a traffic matrix $[T_{i,j}]$. The $(i, j)$ entry, $T_{i,j} = \sum_n f_{i,j}^{(n)}$, contains the total traffic originating from node $i$ that is destined for node $j$, where $f_{i,j}^{(n)}$ is traffic flow $n$ from node $i$ to node $j$. The problem of joint routing and scheduling of the traffic flow in conventional multi-hop networks has been formulated in [18] as an integer linear programming problem. In assigning time slots, two nearby links cannot transmit together if they can mutually interfere with each other. This falls within the framework of a coloring problem.

With PNC, the coloring problem takes on a new angle: the traffic of a PNC flow at alternate links must take on the same color (same time slots). In addition, as far as PNC is concerned, the individual make-ups of the flows between node $i$ and $j$, $f_{i,j}^{(n)}$, is not important. It is the aggregate traffic $T_{i,j}$ that matters. Also, it is conceivable that PNC can also be used for uni-directional individual flows as long as there is bidirectionality for the aggregate flow. That is, the amount of bidirectional traffic at the "aggregate" level is $\min(T_{i,j}, T_{j,i})$ and they can leverage PNC. The rest, $\max(T_{i,j}, T_{j,i}) - \min(T_{i,j}, T_{j,i})$, may use the conventional scheme. We believe routing and resource allocation in PNC is a topic of much interest for more in-depth future research.

### B. Flow Decomposition

Due to various reasons, including interference and synchronization, some of the nodes on the flow path can leverage PNC while others cannot. In general, an end-to-end path may need to be decomposed into several paths, some using PNC while other using the conventional scheme. With such decomposition, a flow essentially becomes a sequence of sub-flows. Fig. 9 depicts an example of decomposition of a flow into three sub-flows, where PNC is used by sub-flow1 and sub-flow3, and the conventional scheme is used by sub-flow2. With respect to the resource allocation problem mentioned in Part A, the decomposition will also alter the constraints in the optimization problem.

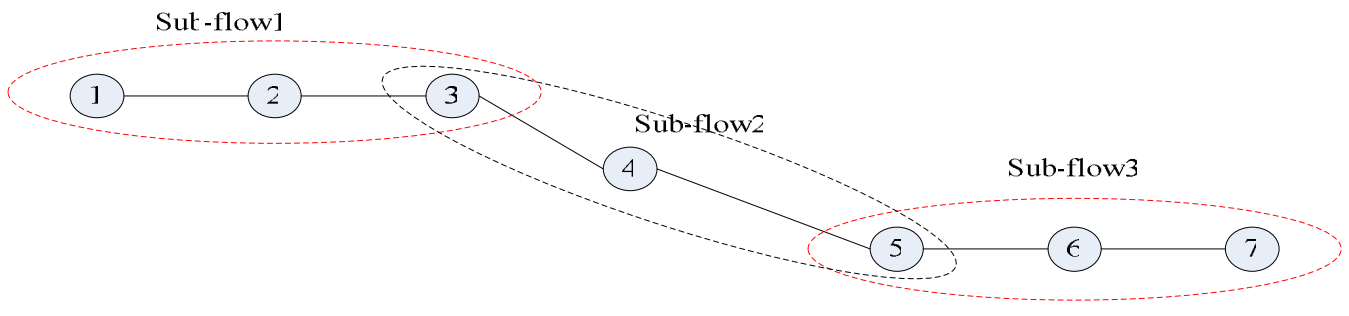

Figure 15. Illustration of flow decomposition

## VII. CONCLUSION

This paper has introduced a novel scheme called *Physical-layer Network Coding* (PNC) that significantly enhances the throughput performance of multi-hop wireless networks. Instead of avoiding interference caused by simultaneous electromagnetic waves transmitted from multiple sources, PNC embraces interference to effect network-coding operation directly from physical-layer signal modulation and demodulation. With PNC, signal scrambling due to interference, which causes packet collisions in the MAC layer protocol of traditional wireless networks (e.g., IEEE 802.11), can be eliminated.

For PNC to be feasible, network-coding arithmetic must be realized with direct electromagnetic-wave mixing, coupled with appropriate modulation and demodulation schemes. This paper has presented the fundamental condition for the equivalence of the conventional network-coding operation and PNC operation. We have illustrated the application of the condition on the 4-PAM signal modulation scheme.

We have shown that PNC can achieve 100% improvement in physical-layer throughput over the traditional multi-hop transmission scheduling scheme, and 50% over the straightforward network coding scheme, in terms of the number of time slots needed to transmit a frame of data (i.e., when errors are ignored). In addition, the throughput achieved by PNC in a regular linear multi-hop network is that of the theoretical upper-bound throughput. Further discussions on synchronization issues are given in the appendices.

We have additionally shows that the information-theoretic capacity a two-way relay channel with PNC is twice that with traditional transmission in the regime where SNR is higher than 0 dB. This is established by the convergence of the analytical lower and upper bounds for the capacity of the PNC. For the low SNR regime, the lower and upper bounds do not converge, and the actual information capacity of PNC remains an open problem.

## APPENDIX 1: SYNCHRONIZATION OF MULTIPLE-NODE PNC CHAIN

It may appear at first glance that synchronization problem of the $N$-node ($N > 3$) case may cause PNC to break down, particularly for large $N$. The goal of this appendix is to examine this issue more carefully. In particular, we argue that the detection scheme in PNC does not break down just because $N$ is large.

We first review prior work on synchronization relevant to the three-node case. PNC requires time, carrier-frequency and carrier-phase synchronizations. Time and carrier-frequency synchronizations have been actively investigated by researchers in the fields of OFDMA, wireless-sensor network, and/or cooperative transmission. In particular, methods for joint estimation of carrier-frequency errors, timing error and channel response [19, 20] have been proposed for OFDMA networks, while reference broadcast synchronization (RBS) [21] and TPSN [22] have been proposed for wireless sensor networks. Carrier-phase synchronization has been studied in the field of coherent cooperation and/or distributed beam forming recently. For example, positive results have been obtained in [23] with a master-slave architecture to prove the feasibility of the distributed beam forming technique. Another carrier-phase and carrier-frequency synchronization scheme has also been proposed in [24] where a beacon is used to measure round trip phase delays between the transmitter and the destination.

The goal of this appendix is not to extend the prior results on the three-node case. We assume the feasibility of synchronization in a three-node chain is a given based on these prior results, and consider how the $N$-node case can make use of 3-node synchronization. A possible approach is to partition the long chain into multiple three-node local groups, as illustrated in Fig. A1.1, and then synchronize them in a successive manner. Suppose the synchronization for three-node can be achieved with reasonable error bounds for phase, frequency, and time (see Appendix 2, where we argue that PNC detection is not very sensitive to synchronization errors), represented by, say, $\theta,\ 2\Delta f,\ \Delta t$ for consistency with the notation in Appendix 2. An issue is the impact of these errors on the *N*-node chain.

For *N*-node synchronization, let us divide the time into two parts: the synchronization phase and the data-transmission phase, as shown in Fig. A2.2. These two phases are repeated periodically, say once every, $T_P$ seconds. The synchronization phase lasts $T_S$ seconds and the data transmission phase lasts $T_D$ seconds, with $T_S + T_D = T_P$ . The PNC data transmission described in the text comes into play only during the data-transmission phase. The synchronization overhead is $T_S / T_P$, with $T_S$ depending on the synchronization handshake overhead, and $T_P$ depending on the speed at which the synchronizations drift as time progresses. That is, the faster the drift, the smaller the $T_P$, because one will then need to perform resynchronization more often. It turns out that the *N*-node case increases the $T_S$ required, but not the $1/T_P$ required as compared to the 3-node case, as detailed below.

For the *N*-node chain, let us divide the synchronization phases into two subphases. The first subphase is responsible for synchronizing all the odd-numbered nodes and the second for all the even-numbered nodes. We describe only subphase 1 here (phase 2 is similar). With reference to Fig. A1.1, we divide the *N* nodes into $M = \left\lfloor \frac{N-1}{2} \right\rfloor$ basic groups (BGs) and denote them by BG *j*, where *j* is index of the BGs. Let $\Delta t_{BG}$ be the time needed to synchronizing the two odd nodes in one BG (using, say, one of the prior methods proposed by others). Consider BG1. Let us assume that it is always the case that the right node (in this case, node 3) attempts to synchronize to the left node (in this case, node 1). After this synchronization, the phase, frequency and time errors between nodes 1 and 3 are $\theta,\ 2\Delta f,\ \Delta t$ . In the next $\Delta t_{BG}$ time, we then synchronizes node 5 to node 3 in BG2. So, a total of time of $M\Delta t_{BG}$ are needed in subphase 1. Including subphase 2, $T_S = (N-2)\Delta t_{BG}$ .

It turns out that with a cleverer scheme, subphase 2 can be eliminated and $T_S$ can be reduced roughly by half. But that is not the main point we are trying to make here. The main issue is that with the above method, the bounds of the synchronization errors of node *N* with respect to node 1 become $M\theta,\ 2M\Delta f,\ M\Delta t$ and these errors grow in an uncontained manner as *N* increases! Will PNC therefore break down as *N* increases?

Recall that for PNC detection, a receiver receives signals simultaneously from only the two adjacent nodes. For example, say, *N* is odd. The reception at node 2 depends only on the synchronization between nodes 1 and 3; and the reception at node *N*-1 only depends on the synchronization of nodes *N*-2 and *N*. In particular, it is immaterial that there is a large synchronization error between nodes 1 and *N*. So, the fact that the end-to-end synchronization errors have grown to $M\theta,\ 2M\Delta f,\ M\Delta t$ is not important. Only the local synchronization errors, $\theta,\ 2\Delta f,\ \Delta t$ , are important. The same reasoning also leads us to conclude that how often synchronization should be performed (i.e., $1/T_P$ ) does not increase with *N* either, since it is only the drift within 3 nodes that are important as far as PNC detection is concerned.

Of course, $T_S$ grows with *N*, but only linearly. If $\Delta t_{BG}$ is small compared with $T_P$, this is not a major concern. In practice, however, we may still want to impose a limit on the chain size *N* not just to limit the overhead $T_S$, but also for other practical considerations, such as routing complexities, network management, etc.

Appendix 2 examines the impact of synchronization errors on PNC, and discusses what if synchronization is not performed at all (or very rarely).

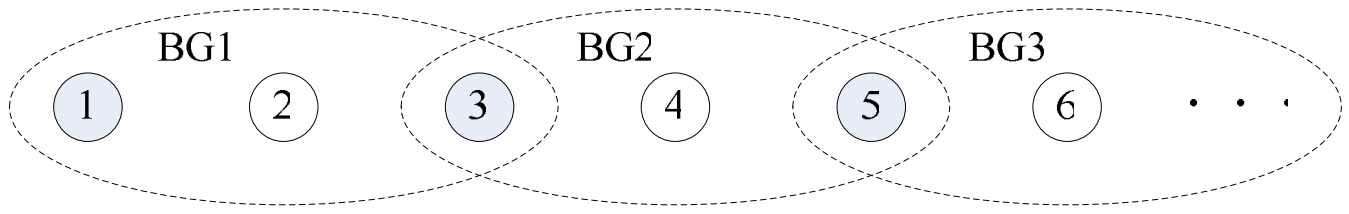

Figure A1.1. Synchronization for multiple nodes

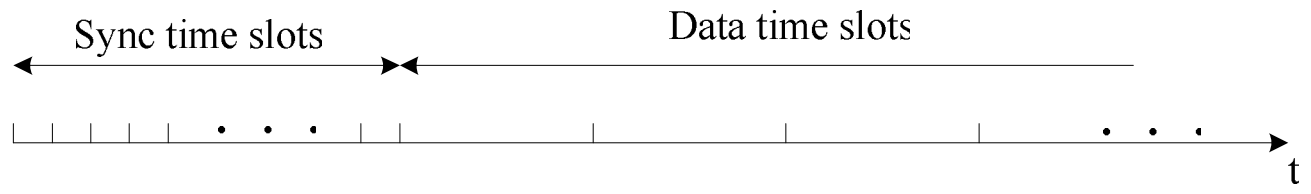

Figure A1.2. Partitioning of time into synchronization phase and data-transmission phase.

## APPENDIX 2: PERFORMANCE PENALTY OF SYNCHRONIZATION ERRORS

In this appendix, we investigate the performance penalty of synchronization errors on PNC. This framework is applicable to situations where synchronization is not perfect (e.g., with respect to Appendix 1, synchronization may become imperfect in between two synchronization periods) as well as where synchronization is not performed at all.

*1) Penalty of carrier-phase synchronization errors:*

We first consider carrier-phase errors. We assume that the relative carrier-phase offset of the two input signals are known to the receiver[1]. Consider BPSK as an example. The two received signals can be written as:

$$s_1(t) = a_1 \cos(2\pi f t)$$

$$s_2(t) = a_2 \cos(2\pi f t + \theta)$$

where $a_1$ and $a_2$ are the information bits,

[1] Before the adjacent transmitters transmit their data concurrently as per PNC, they could first take turn transmitting a preamble in a non-overlapping manner. The receiver can then derive the phase difference from the two preambles. Frequency and time offsets can be similarly determined using preambles. Note that this is different from synchronization, since the transmitters do not adjust their phase, frequency and time differences thereafter. The receivers in a PNC chain simply accept the synchronization errors the way they are.

$\theta$ $(-\pi/2 \le \theta < \pi/2)$ is the phase offset and $f$ is the carrier frequency. Note that we only need to deal with the case when $-\pi/2 \le \theta < \pi/2$ . If $\pi/2 < \theta \le 3\pi/2$ , we can simply substitute $a_2$ with $a_2' = -a_2$, and $\theta$ with $\theta' = \theta - \pi$.

Suppose that the receiver positions the phase of its mixing signal at $\theta/2$. Then, the baseband signal recovered can be written as

$$r(\theta) = r_1(\theta) + r_2(\theta)$$
$$= \int_0^T a_1 \cos(2\pi ft)\cos(2\pi ft + \theta/2)dt +$$
$$\int_0^T a_2 \cos(2\pi ft + \theta)\cos(2\pi ft + \theta/2)dt$$
$$= a_1 T\cos(\theta/2)/2 + a_2 T\cos(\theta/2)/2$$

We see that the phase error causes a decrease in the received signal power. The power penalty is

$$\Delta\gamma(\theta) = r^2/(a_1T/2 + a_2T/2)^2 = \cos^2(\theta/2)$$

If the phase offset is distributed uniformly over $[-\pi/2, \pi/2]$ (this is a reasonable assumption if synchronization is not performed at all), the average power penalty is

$$\overline{\Delta\gamma(\theta)} = \frac{1}{\pi}\int_{-\pi/2}^{\pi/2} \cos^2(\theta/2)d\theta = \frac{1}{\pi} + \frac{1}{2} = -0.87dB$$

That is, even if carrier-phase synchronization is not performed, the average SNR penalty is less than 1 dB.

In the worst case, the power penalty is $\Delta\gamma(\pi/2) = -3dB$, which is still generally acceptable in the wireless environment. To avoid the worst-case penalty and to obtain the average power penalty performance, the transmitters could intentionally change their phases from symbol to symbol using a “phase increment” sequence known to the receivers. If the phase-increment sequences of the two transmitters are not correlated, then certain symbols are received with low error rates and certain symbols are received with high error rates during a data packet transmission. With FEC coding, the overall packet error rate can be reduced. This essentially translates the power penalty to data-rate penalty.

*2) Penalty of carrier frequency synchronization errors:*

For the analysis of frequency-synchronization errors, suppose that the two signals are

$$s_1(t) = a_1 \cos(2\pi(f - \Delta f)t)$$
$$s_2(t) = a_2 \cos(2\pi(f + \Delta f)t)$$

where $2\Delta f$ is the carrier frequency offset. Let us assume $\Delta f \cdot T \ll 1$. The receiver sets the frequency of its mixing signal to $f$ . The recovered baseband signal is

$$r(\Delta f) = r_1(\Delta f) + r_2(\Delta f)$$
$$= \int_0^T a_1 \cos(2\pi(f - \Delta f)t)\cos(2\pi ft)dt +$$
$$\int_0^T a_2 \cos(2\pi(f - \Delta f)t)\cos(2\pi ft)dt$$
$$= a_1 \sin(2\pi\Delta fT)/4\pi\Delta f + a_2 \sin(2\pi\Delta fT)/4\pi\Delta f$$

The power penalty is

$$\Delta\gamma(\Delta f) = r^2/(a_1T/2 + a_2T/2)^2 = \sin^2(2\pi\Delta fT)/(2\pi\Delta fT)^2$$

In Fig. A2.1, we plot $\Delta\gamma$ against $\Delta f \cdot T$ for $0 \le \Delta f \cdot T \le 0.1$. It can be seen that the maximum power penalty is less than 0.6dB.

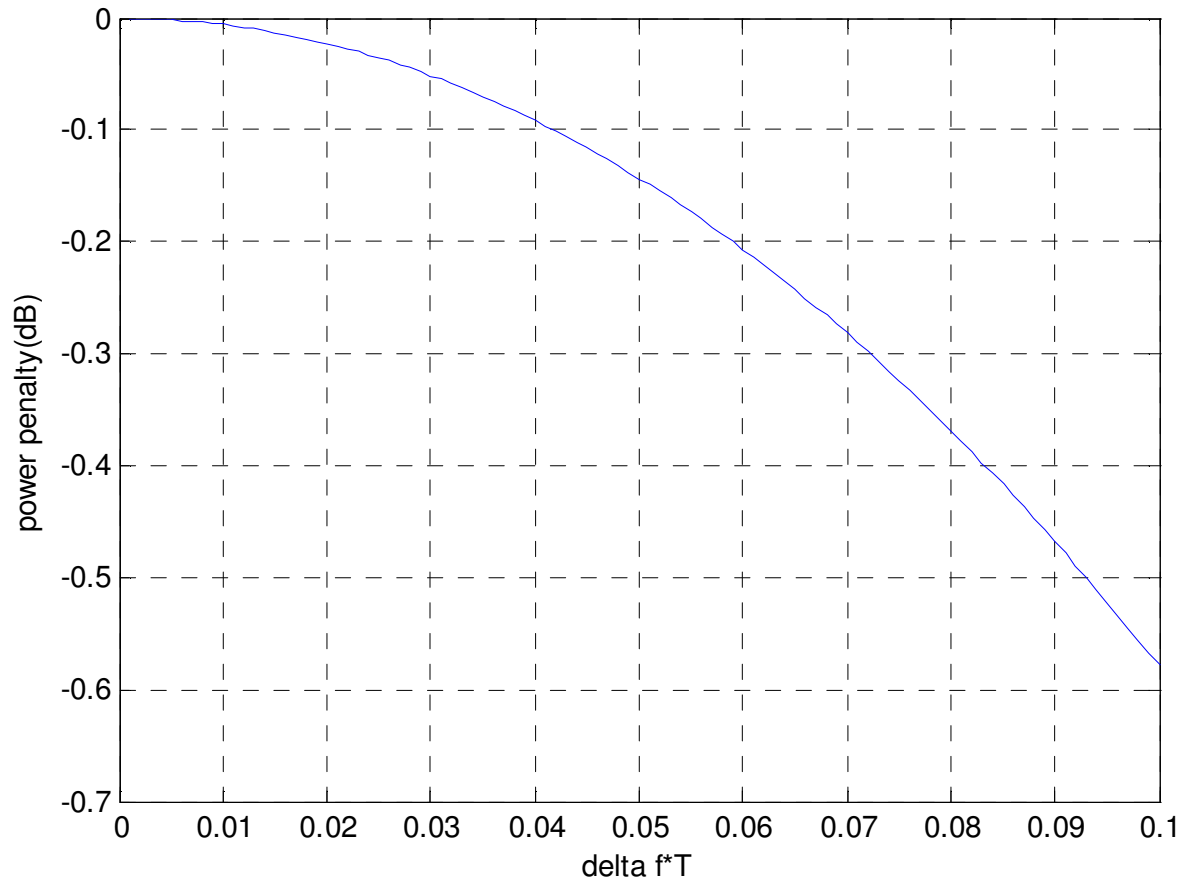

Figure A2_1. Power penalty of frequency synchronization errors

*3) Penalty of time synchronization errors:*

Ref. [25] analyzes the impact of time synchronization errors on the performance of cooperative MISO systems, and show that the clock jitters as large as 10% of the bit period actually do not have much negative impact on the BER performance of the system. Based on the similar methodology, we can also analyze the impact of time synchronization error toward the performance of PNC.

Let $\Delta t$ be the time offset of the two input signals. The two transmitted signals can be written as:

$$s_1(t) = \sum_{l=-\infty}^{\infty} a_1[l]\cos(2\pi ft)g(t - lT)$$
$$s_2(t) = \sum_{l=-\infty}^{\infty} a_2[l]\cos(2\pi ft)g(t - lT - \Delta t)$$

where, $a_j[l]$ is the $l^{th}$ bit of the signal $s_j(t)$, and $g(t)$ is pulse. The baseband signal can be written as

$$r(t) = r_1(t) + r_2(t)$$
$$= \frac{1}{2}\sum_l a_1[l]g(t - lT) + a_2[l]g(t - lT - \Delta t)$$

After the match filter, the receiver samples the signal at time instances $t = kT - \Delta t/2$ (i.e., at the middle of the offset). We

then have

$$r[k] = r_1[k] + r_2[k]$$
$$= \frac{1}{2}\sum_l \{a_1[l]p((k-l)T + \Delta t/2) + a_2[l]p((k-l)T - \Delta t/2)\}$$
$$= (a_1[k] + a_2[k])p(\Delta t/2)/2 +$$
$$\frac{1}{2}\sum_{l,l\neq k} \{a_1[l]g((k-l)T + \Delta t/2) + a_2[l]p((k-l)T - \Delta t/2)\}$$

where, $p(t)$ is the response of the receiving filter to the input pulse $g(t)$. As widely used in practice, the raised cosine pulse shaping function, $p(t) = \frac{\sin(\pi t/T)\cos(\pi\beta t/T)}{\pi t/T\bullet(1-4\beta^2 t^2/T^2)}$, is chosen. We see that the time synchronization errors not only decrease the desired signal power, but also introduce inter-symbol interference (ISI). Therefore, we use SINR (signal over noise and interference ratio) penalty here to evaluate the performance degradation. The SINR penalty can be calculated as

$$\Delta\gamma(\Delta t) = SINR(\Delta t) - SNR_0$$
$$= 10\log_{10}(p(\Delta t/2))^2 - 10\log_{10}(\frac{\sigma_{isi}^2 + \sigma_n^2}{\sigma_n^2})$$

where

$$\sigma_{isi}^2 = E\{(\sum_{l,l\neq k} a_1[l]p((k-l)T + \Delta t/2) + a_2[l]p((k-l)T - \Delta t/2))^2\}$$

is the variance of the inter-symbol interference. Figure A2.2 plots the power penalty versus $\Delta t/T$, where the $SNR_0$ is set to 10dB and the roll factor of the raised cosine function is set to 0.5. The worst-case SINR penalty is about -2.2 dB. If we assume the time synchronization error to uniformly distribute over [-*T*/2, *T*/2], we can calculate the average SINR penalty as:

$$\overline{\Delta\gamma} = \int_{-.05}^{0.5} \Delta\gamma(\tau)d\tau$$
$$= \int_{-.05}^{0.5} SINR(\tau)d\tau - SNR_0$$
$$= -1.57dB$$

Based on the discussion in this appendix, we can conclude that the performance degradation of 1 to 3dB due to various synchronization errors (including large synchronization errors in the case where a synchronization mechanism is not used at all) is acceptable given the more than 100% throughput improvement obtained by PNC. The discussion has been based on specific examples. More general treatments await further research.

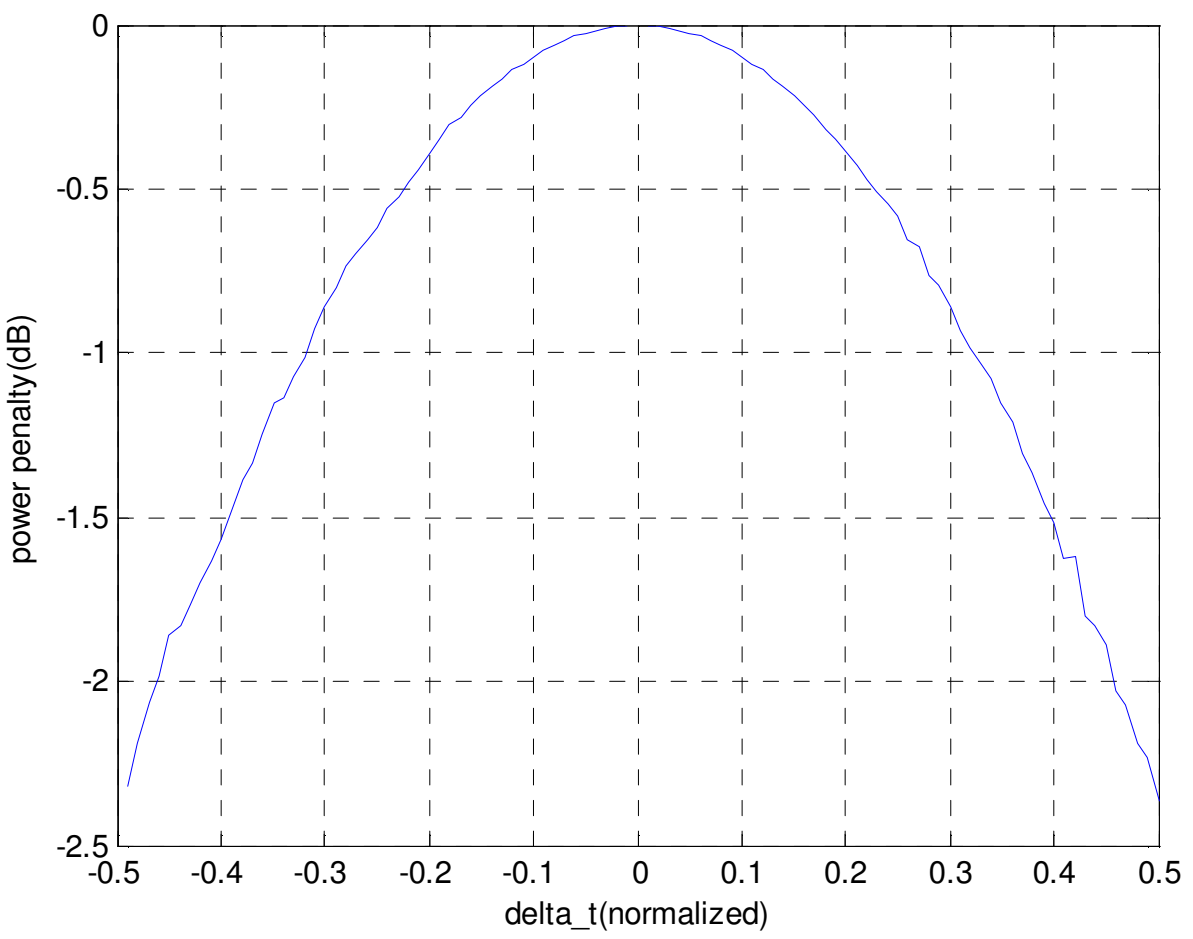

Figure A2.2. Power penalty of time synchronization errors

**First A. Author** (M'76–SM'81–F'87) and the other authors may include biographies at the end of regular papers. Biographies are often not included in conference-related papers. This author became a Member (M) of IEEE in 1976, a Senior Member (SM) in 1981, and a Fellow (F) in 1987. The first paragraph may contain a place and/or date of birth (list place, then date). Next, the author's educational background is listed. The degrees should be listed with type of degree in what field, which institution, city, state or country, and year degree was earned. The author's major field of study should be lower-cased.

The second paragraph uses the pronoun of the person (he or she) and not the author's last name. It lists military and work experience, including summer and fellowship jobs. Job titles are capitalized. The current job must have a location; previous positions may be listed without one. Information concerning previous publications may be included. Try not to list more than three books or published articles. The format for listing publishers of a book within the biography is: title of book (city, state: publisher name, year) similar to a reference. Current and previous research interests ends the paragraph.

The third paragraph begins with the author's title and last name (e.g., Dr. Smith, Prof. Jones, Mr. Kajor, Ms. Hunter). List any memberships in professional societies other than the IEEE. Finally, list any awards and work for IEEE committees and publications. If a photograph is provided, the biography will be indented around it. The photograph is placed at the top left of the biography. Personal hobbies will be deleted from the biography.